\documentclass[10pt,twocolumn,superscriptaddress,english,
prl,
floatfix,aps]{revtex4-2}
\usepackage[T1]{fontenc}           
\usepackage[utf8]{inputenc}
\usepackage{amsmath}
\usepackage{amssymb}
\usepackage{graphicx}
\usepackage{xspace}
\usepackage{physics}
\usepackage{verbatim}

\usepackage[
separate-uncertainty  = true,
uncertainty-separator =  {\pm},
exponent-product  =  \cdot,
quotient-mode  =  fraction
]{siunitx}
\usepackage{lipsum}
\usepackage[backref=none,bookmarksnumbered=true,bookmarks=true,
	bookmarksopen=true,colorlinks=true,citecolor=blue,linkcolor=blue,
	anchorcolor=green,urlcolor=blue,unicode=false]{hyperref}

\makeatletter
\@ifundefined{textcolor}{}
{%
 \definecolor{BLACK}{gray}{0}
 \definecolor{WHITE}{gray}{1}
 \definecolor{RED}{rgb}{1,0,0}
 \definecolor{GREEN}{rgb}{0,1,0}
 \definecolor{BLUE}{rgb}{0,0,1}
 \definecolor{CYAN}{cmyk}{1,0,0,0}
 \definecolor{MAGENTA}{cmyk}{0,1,0,0}
 \definecolor{YELLOW}{cmyk}{0,0,1,0}
}

\usepackage{soul}
\usepackage{braket}
\usepackage[backref=none,
bookmarksnumbered=true,
bookmarks=true,
bookmarksopen=true,
colorlinks=true,
citecolor=blue,
linkcolor=blue,
anchorcolor=green,
urlcolor=blue,unicode=false]{hyperref}

\let\baraccent=\= 
\renewcommand{\=}[1]{\stackrel{#1}{=}} 

\newcommand{\didv}{\ensuremath{\mathrm{d}I/\mathrm{d}V}\xspace}

\newcommand{\nbse}{NbSe$_2$}
\newcommand{\spat}{$\SI{100}{\pico\ampere},\SI{10}{\milli\volt}$}
\newcommand{\sptop}{$\SI{500}{\milli\volt},\SI{80}{\pico\ampere}$}

\makeatother

\usepackage{babel}

\begin{document}

\title{Quantum spins and hybridization in artificially-constructed chains of magnetic adatoms on a superconductor}

\author{Eva Liebhaber}
\affiliation{Fachbereich Physik, Freie Universit\"at Berlin, 14195 Berlin, Germany}

\author{Lisa M. R\"{u}tten}
\affiliation{Fachbereich Physik, Freie Universit\"at Berlin, 14195 Berlin, Germany}

\author{Ga\"el Reecht}
\affiliation{Fachbereich Physik, Freie Universit\"at Berlin, 14195 Berlin, Germany}

\author{Jacob F. Steiner}
\affiliation{Dahlem Center for Complex Quantum Systems and Fachbereich Physik, Freie Universit\"at Berlin, 14195 Berlin, Germany}

\author{Sebastian Rohlf}
\affiliation{Institut für Experimentelle und Angewandte Physik, Christian-Albrechts-Universit\"at zu Kiel, 24118 Kiel, Germany}

\author{Kai Rossnagel}
\affiliation{Institut für Experimentelle und Angewandte Physik, Christian-Albrechts-Universit\"at zu Kiel, 24118 Kiel, Germany}
\affiliation{Ruprecht Haensel Laboratory, Deutsches Elektronen-Synchrotron DESY, 22607 Hamburg, Germany}

\author{Felix von Oppen}
\affiliation{Dahlem Center for Complex Quantum Systems and Fachbereich Physik, Freie Universit\"at Berlin, 14195 Berlin, Germany}

\author{Katharina J. Franke}
\email{franke@physik.fu-berlin.de}
\affiliation{Fachbereich Physik, Freie Universit\"at Berlin, 14195 Berlin, Germany}

%

\begin{abstract}
Magnetic adatom chains on surfaces constitute fascinating quantum spin systems. Superconducting substrates suppress interactions with bulk electronic excitations but couple the adatom spins to a chain of subgap Yu-Shiba-Rusinov (YSR) quasiparticles. Using a scanning tunneling microscope, we investigate such correlated spin-fermion systems by constructing Fe chains adatom by adatom on superconducting \nbse. The adatoms couple entirely via the substrate, retaining their quantum spin nature. In dimers, we observe that the deepest YSR state undergoes a quantum phase transition due to Ruderman-Kittel-Kasuya-Yosida interactions, a distinct signature of quantum spins. Chains exhibit coherent hybridization and band formation of the YSR excitations, indicating ferromagnetic coupling. Longer chains develop separate domains due to coexisting charge-density-wave order of \nbse. Despite the spin-orbit-coupled substrate, we find no signatures of Majoranas, possibly because quantum spins reduce the parameter range for topological superconductivity. We suggest that adatom chains are versatile systems for investigating correlated-electron physics and its interplay with topological superconductivity.
\end{abstract}

\pacs{%
			} 
\maketitle 


\section{Introduction}
Ever since the early days of quantum mechanics, chains of coupled quantum spins have been a paradigm of strongly interacting quantum systems \cite{Bethe1931}. Magnetic adatoms on surfaces have been an attractive platform for realizing assemblies of interacting spins, also because atom manipulation allows for designing a wide variety of structures and couplings \cite{Choi2019}. On normal-metal substrates, the adatom spins couple to and relax via the particle-hole continuum of the conduction electrons. This relaxation can be suppressed on superconducting substrates as a consequence of their excitation gap \cite{Heinrich2013b}. The gap effectively decouples the adatom chain from the bulk electronic excitations of the substrate, while retaining the spin-spin coupling due to 
the Ruderman-Kittel-Kasuya-Yosida (RKKY) \cite{Ruderman1954, Kasuya1956, Yosida1957} and Dzyaloshinsky-Moriya interactions \cite{Dzyaloshinsky1958,Moriya1960}, which are mediated by virtual higher-energy states of the quasiparticle continuum.

Despite the decoupling from bulk excitations of the substrate, chains of magnetic adatoms remain intertwined with localized fermionic subgap excitations. Exchange coupling an adatom spin to the substrate electrons induces discrete Bogoliubov quasiparticles within the superconducting gap known as Yu-Shiba-Rusinov (YSR) excitations \cite{Yu1965,Shiba1968,Rusinov1969,Balatsky2006}. Due to the exchange coupling, the adatom binds localized quasiparticles which (partially) screen the adatom spin. Chains of magnetic adatoms on superconductors are thus intriguing many-body systems, which couple quantum spins to an associated chain of well-defined localized fermionic quasiparticles \cite{Steiner2021}. Such coupled spin-fermion systems are paradigmatic for correlated-electron physics, underlying the physics of Kondo lattice systems \cite{Tsunetsugu1997} and of high-$T_c$ superconductors \cite{Lee2006}. Unlike these settings, magnetic adatoms on superconductors provide a coupled spin-fermion system which is amenable to bottom-up design in a multitude of geometries.

Chains of magnetic adatoms on superconductors are highly versatile as a result of quantum phase transitions between states with different numbers of bound quasiparticles \cite{Sakurai1970,Zitko2011,Oppen2021}. These transitions arise from the competition between superconducting pairing and the exchange coupling between adatom spin and substrate \cite{Sakurai1970,Franke2011,Farinacci2018,Malavolti2018,Huang2020}, and are associated with discrete changes in the effective adatom spin due to Kondo-like screening. In chains of magnetic adatoms, different screening states have different spin-spin interaction energies, leading to rich magnetic phase diagrams as a function of the binding energy of screening YSR quasiparticles and spin-spin interactions \cite{Steiner2021}. The competition between screening and spin-spin interactions is specific to quantum spins and absent in models based on classical spins. Intriguingly, the magnetic phase diagrams can be probed directly by tunneling experiments at subgap bias voltages. Tunneling locally binds or unbinds a quasiparticle, introducing a sudden change of the local effective adatom spin. The response to this local perturbation provides signatures in the tunneling spectra which are characteristic of the magnetic ordering \cite{Steiner2021}.

Here, we report experiments on magnetic chains built Fe adatom by Fe adatom on a $2H$-\nbse\ substrate, a layered superconductor. Due to the quasi two-dimensional structure, the YSR resonances have a larger spatial extent \cite{Menard2015,Kezilebieke2018}, which facilitates coupling  along the chain even for well-separated Fe adatoms. We track the spatially resolved excitation spectra all the way from the monomer to longer chains. Quantum phase transitions induced by the RKKY coupling reveal the intimate coupling of the quantum spins of the adatoms to the fermionic subgap quasiparticles. The YSR excitations hybridize coherently along the chain and the resulting formation of band-like subgap excitations with increasing chain length suggests ferromagnetic coupling between adatom spins. Longer chains are affected by the charge-density-wave (CDW) order of the $2H$-\nbse\ substrate that coexists with superconductivity.

Previous studies of magnetic chains on superconducting substrates were motivated by the search for topological superconductivity and Majorana bound states. These experiments  relied on densely-packed adatoms, formed either by self assembly \cite{NadjPerge2014,Ruby2015chains, Feldman2016, Pawlak2016, Ruby2017} or by atom manipulation \cite{Kim2018, Schneider2020, Mier2021, Schneider2021a,Schneider2021b}. The theoretical description of these chains assumes classical spin structures and bands of hybridizing adatom $d$ orbitals, which cross the Fermi level \cite{NadjPerge2013,NadjPerge2014,Li2014a,Peng2015}. Importantly, the chains investigated here are in a very different regime, previously studied only for YSR dimers \cite{Rusinov1969, Yao2014a, Ruby2018, Choi2018, Beck2020, Ding2021, Kamlapure2019, Kuester2021}. We construct magnetic adatom chains in the dilute limit \cite{Pientka2013}, in which the adatoms are spaced sufficiently far that direct hybridization of their $d$ orbitals is negligible. The adatoms couple only via spin-spin interactions mediated by the substrate as well as hybridization of their YSR excitations. In this limit, subgap bands emerge from correlated spin-fermion dynamics and are in general not of single-particle nature. 

Dilute chains of magnetic adatoms have also been predicted to host topological superconductivity and Majorana end states \cite{Pientka2013,Poyhonen2014,Pientka2015,Pawlak2019,Steiner2021}. These studies assumed classical spin textures, neglecting the quantum nature of the adatom spin. Quantum spins, however, can have dramatically reduced parameter space exhibiting topological superconductivity \cite{Steiner2021}, and we do not observe signatures of Majorana end states in any of our chains. In general, one can realize a wide variety of quantum spin chains by tuning the effective adatom spins, the magnetic anisotropy, or the sign and magnitude of the RKKY coupling. 
While higher-spin adatoms promise more extensive topological superconducting phases, lower-spin adatoms promise more pronounced signatures of correlated electron physics.
This makes dilute adatom chains exciting model systems for correlated
electron physics, for topological superconductivity, and for their interplay.

\section{Results}


\begin{figure}[h]\centering
\includegraphics[width=0.95\linewidth]{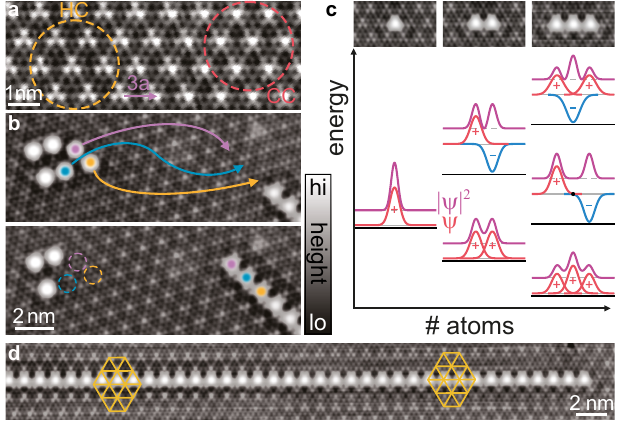}
\caption{\textbf{Assembly of Fe chains on \nbse .}
\textbf{a} Atomic-resolution topography of a clean \nbse\ surface. Regions with HC (CC) CDW patterns are indicated in yellow (red). \textbf{b} Topography images illustrating the sequence of atomic-manipulation steps to obtain an Fe chain. \textbf{c} STM images of Fe monomer, dimer, and trimer along with a schematic illustration of YSR hybridization. YSR wave functions of the monomers (taken as Gaussian shaped) with red and blue indicating positive and negative amplitude in overall hybridized wave function $\psi$ (with $\left|\psi\right|^2$ shown in purple). A tight-binding-like hybridization into linear combinations of monomer wave functions predicts characteristic maxima and nodal planes of the YSR wave function. \textbf{d} Topography of a long chain. The yellow grids illustrate the position of the CDW maxima relative to the atoms of the chain. Constant-current set point for all topographies $\SI{100}{\pico\ampere},\SI{10}{\milli\volt}$.
}
\label{Fig:figure1}
\end{figure}

We build chains on the van der Waals-layered superconductor $2H$-\nbse\, starting with a single Fe atom and continuing all the way to a 51-atom chain. The \nbse\, substrate forms a CDW \cite{Malliakas2013, Dai2014, Arguello2014} which is reflected in periodic height modulations superimposed on the atomic corrugation (see STM image in Fig.\,\ref{Fig:figure1}a). The CDW is weakly incommensurate with a period of $\sim 3a$ (here, $a$ is the lattice constant of \nbse), so that the CDW shifts slowly relative to the atomic lattice. When the CDW maximum is centered on the hollow site (hollow centered, HC), the topography exhibits a three-petaled shape (yellow area in Fig.\,\ref{Fig:figure1}a), while the Se-atom centered configuration (chalcogen centered, CC) exhibits a petalless pattern (red area). Superconductivity appears below a critical temperature of $\SI{7.2}{\kelvin}$.  The anisotropic superconducting gap of about $\SI{1}{\milli\electronvolt}$ results in two-humped coherence peaks in the \didv spectra (see gray spectrum in Fig.\,\ref{Fig:figure2}a) \cite{Rahn2012,Yokoya2001,Noat2010,Guillamon2008}. 

As deposited, Fe atoms adsorb in the two distinct hollow sites of the terminating Se layer, with or without Nb atom underneath (metal centered and hollow centered, respectively). 
The two kinds of Fe adatoms exhibit distinct YSR spectra, reflecting different crystal-field splittings as well as exchange and potential scatterings \cite{Liebhaber2020}. To maximize YSR hybridization, we assemble shorter chains such that all Fe atoms are positioned on identical (hollow-centered) sites (Fig.\,\ref{Fig:figure1}b-c). 

For individual Fe atoms, local variations associated with the CDW shift the energy and modify the wave functions of the YSR states \cite{Liebhaber2020}. Consequently, effective hybridization of YSR states requires not only that the adatoms are located at identical adsorption sites of the atomic lattice, but also at comparable positions with respect to the CDW. For this reason, we place hollow-centered Fe adatoms on a sequence of maxima of the CDW, resulting in a dilute chain with an adatom distance of $3a$. We show that with this choice, the YSR excitations of adjacent Fe atoms are sufficiently similar that they hybridize into symmetric and antisymmetric linear combinations (see sketch in Fig.\,\ref{Fig:figure1}c). Successively placing Fe atoms at a distance of $3a$ (Fig.\,\ref{Fig:figure1}b), we find that chains with up to about ten adatoms remain essentially uniform. Longer chains become inhomogeneous due to the incommensurate nature of the CDW (Fig.\,\ref{Fig:figure1}d).

\subsection{Monomer}

We begin with an analysis of the monomer. YSR states have been observed in \nbse\ for magnetic atoms located in the bulk \cite{Menard2015, Senkpiel2018} and on the surface \cite{Kezilebieke2018, Liebhaber2020, Yang2020}. A differential conductance (\didv) spectrum of a Fe monomer at the specified adsorption site and recorded with a superconducting Nb tip is shown in Fig.\,\ref{Fig:figure2}a (black line; tip gap indicated in gray). 
We resolve four pairs of subgap YSR resonances labeled $\alpha$-$\delta$ at symmetric bias voltages. These emerge from exchange coupling four half-filled $d$ orbitals to corresponding conduction electron channels \cite{Ruby2016,Oppen2021}, indicating that the Fe atom is in a spin-2 state. 

Due to the Kondo-like screening, the effective spin $S_\mathrm{eff}$ of the monomer depends on the number $Q$ of bound quasiparticles. Depending on the strength of the exchange coupling within the channel, each channel can bind a quasiparticle or not, with the two ground states separated by a quantum phase transition (see also Supplementary Note 1). Each bound quasiparticle partially screens the bare impurity spin $S=2$. We can deduce the screening state of the four channels by tracking how the energies of the YSR states shift as a function of the adsorption site relative to the CDW \cite{Liebhaber2020}. Our analysis indicates that for Fe on the maximum of the CDW, three YSR resonances ($\alpha$-$\gamma$) are in the screened state. For the fourth resonance ($\delta$), the assignment is less definitive, with the unscreened state being more likely and assumed in the following. With three channels binding a quasiparticle at the impurity site, the spin of the monomer is effectively reduced to $S_\mathrm{eff}=\frac{1}{2}$, and the monomer is in a $(Q=3,S_\mathrm{eff}=\frac{1}{2})$ ground state. (Indications against the less likely assignment of four screened channels are provided in the Supplementary Note 2).

The small effective spin $S_\mathrm{eff}=\frac{1}{2}$ of the adatoms indicates that the spin needs to be treated quantum mechanically. Direct evidence for the quantum spin nature is provided by a Kondo resonance which the Fe atoms induce, when the substrate is in the normal state (see Supplementary Note 4 and Supplementary Fig.\,4). We will see below that a description in terms of quantum spins is also strongly suggested by the spectra of adatom dimers on the superconducting substrate. 

We can obtain fingerprints of the YSR resonances through their wave-function patterns, which will allow us to track how YSR resonances shift and hybridize as adatoms are assembled into chains. As a result of the quantum phase transitions, one must take some care when interpreting YSR patterns. In the unscreened state (no bound quasiparticle), the electron-like wave function is observed at positive bias voltage and the hole-like wave function at negative biases. In contrast, the assignments are reversed in the screened state. Theoretically, the screened ($E_\mathrm{YSR}<0$) and unscreened ($E_\mathrm{YSR}>0$) states are characterized by different signs of the YSR energies, and we use this representation when interpreting our data, see, e.g., Fig.\ \ref{Fig:figure3}(e).

Throughout this paper, we plot the hole-like YSR wave function. (The maps of the electron-like components are shown in the Supplementary Figs.\,7-12.) Specifically, we focus on the YSR states $\alpha$ and $\beta$, which lie deepest inside the gap and are most easily disentangled from the background. Corresponding spectra taken across the Fe atom along the high-symmetry direction of the substrate display long-range oscillations of the YSR wave functions (color plot in Fig.\,\ref{Fig:figure2}b), which have their highest intensity slightly away from the center of the Fe atom. Differential conductance maps of the $\alpha$ and $\beta$ resonances exhibit overall $D_3$ symmetry of the YSR wave functions (brownish plots in Fig.\,\ref{Fig:figure3}a) reflecting the anisotropic Fermi surface of \nbse , the local crystal field, and the symmetry of the CDW \cite{Liebhaber2020}.

\subsection{Dimer} 

\begin{figure}[htbp]\centering
\includegraphics[width=0.9\linewidth]{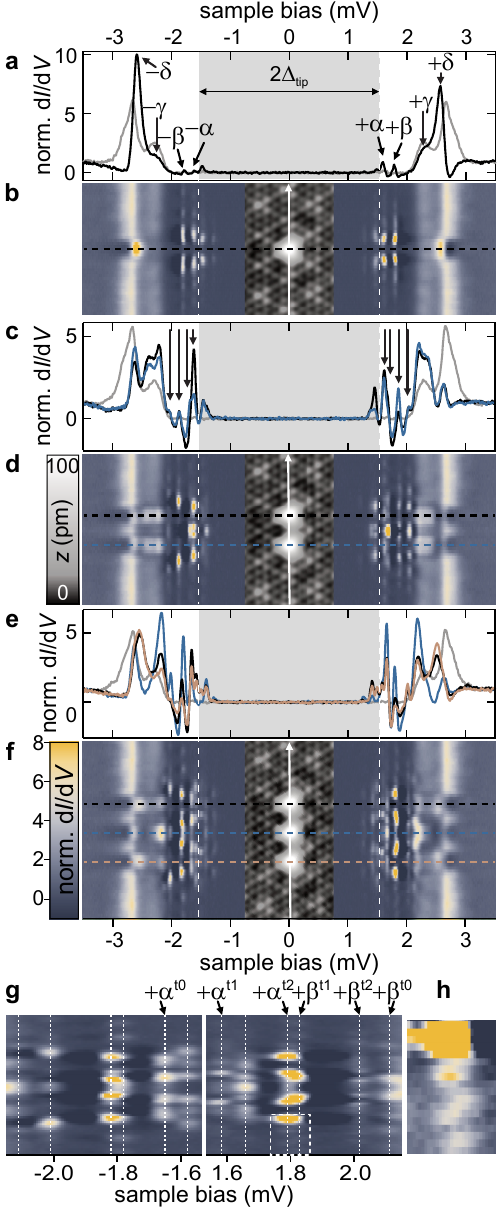}
\caption{\textbf{Evolution of \didv spectra from monomer to trimer. }
\textbf{a,c,e} Normalized differential conductance spectra recorded on the atoms for the monomer \textbf{a}, dimer \textbf{c}, and trimer \textbf{e} and on the bare \nbse\ (gray). Tip gap indicated in gray. Feedback was opened at $\SI{250}{\pico\ampere},\SI{5}{\milli\volt}$ and a modulation of $\SI{15}{\micro\volt}$ was used. In \textbf{a} the $\alpha$-, $\beta$-, $\gamma$- and $\delta$-resonances are indicated by black arrows. In \textbf{c}, the split resonances from $\alpha$ and $\beta$ are indicated by arrows. \textbf{b,d,f} Line profiles recorded along the monomer \textbf{b}, dimer \textbf{d}, and trimer \textbf{f} as illustrated in the topographies (insets; recorded in constant-current mode with set point \spat). Horizontal dashed lines serve as guide to the eye for the precise location of the spectra of \textbf{a,c,e}. Vertical white dashed lines indicate $2\Delta_{\mathrm{tip}}$. \textbf{g} Zoom to subgap range of \textbf{f}. The $\alpha$- and $\beta$-resonances are indicated by the white dashed lines and labeled above the panels. \textbf{h} Close-up of the oscillatory decay around $\SI{1.8}{\milli\volt}$ as indicated by the box in \textbf{g}.
}
\label{Fig:figure2}
\end{figure}
\begin{figure*}[htp]\centering
\includegraphics[width=0.95\linewidth]{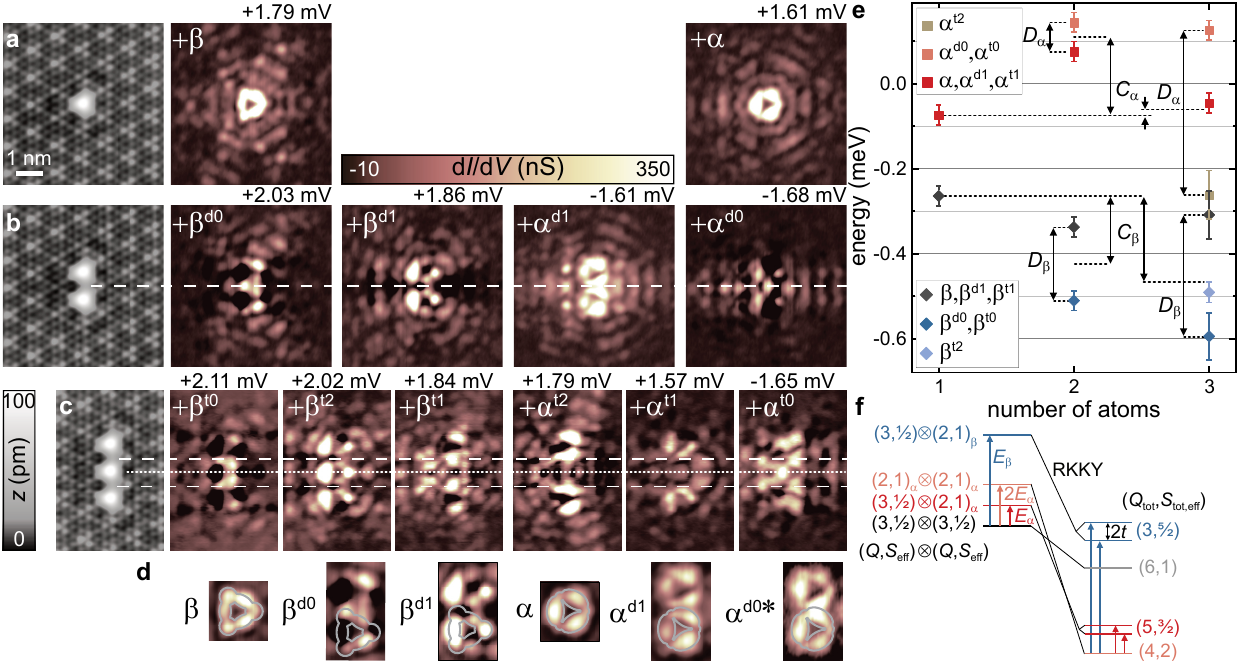}
\caption{\textbf{YSR wave functions of monomer, dimer and trimer. }
\textbf{a-c} STM topographies (constant-current mode with set point \spat) of one to three Fe atoms with spacing of $3a$ in the left. Corresponding constant-contour \didv maps of the (hybridized) YSR states in the monomer (top row), dimer (middle row) and trimer (bottom row). Constant-contour feedback was opened at $\SI{250}{\pico\ampere},\SI{5}{\milli\volt}$ and a modulation of $\SI{15}{\micro\volt}$ was used. Horizontal dashed/dotted lines serve as guide to the eye. \textbf{d} Close-ups around the atoms' center for the monomer and dimer $\alpha$ and $\beta$ states. Gray lines serve as guide to the eye. \textbf{e} Quantitative energy evolution of the $\alpha$ and $\beta$ states upon hybridization. Shift $C$ relative to the monomer and splits $D$ are indicated. Note that the CDW-induced energetic evolution of the monomer YSR states suggests $\alpha$ and $\beta$ to be in the screened-spin regime \cite{Liebhaber2020}, hence, implying a negative ground state energy (of the monomer states). The position of the $\alpha$ and $\beta$ states were determined from deconvolved spectra, which were fit by the appropriate number of Gaussian peaks. The error bars were determined from the standard deviation of the fit, the error margin of the energy gap of the tip, the modulation voltage of the lock-in, and the sampling interval of the data used for the deconvolution, for details see Supplementary Note 6.
\textbf{f} Schematic energy level diagram of the dimer. States of the uncoupled dimer (left) are labeled by $(Q,S_\mathrm{eff})$ for both adatoms. States of the coupled dimer (right) are labeled by the total number of bound quasiparticles and effective spin, $(Q_\mathrm{tot}, S_\mathrm{tot,eff})$. The values of $S_\mathrm{tot,eff}$ assume ferromagnetic RKKY coupling as suggested by band formation in chains (see text). 
}
\label{Fig:figure3}
\end{figure*}

We realize dimers by placing a second hollow-centered Fe atom an adjacent maximum of the CDW at a distance of $3a$. Taking \didv spectra at the centers of the two atoms (black and blue lines in Fig.\,\ref{Fig:figure2}c), we now find two states each in the energy regions of the original $\alpha$ and $\beta$ states. The doubling of the YSR states persists all along the dimer axis with oscillatory intensity distribution and overall mirror symmetry about the center of the dimer (Fig.\,\ref{Fig:figure2}d). This doubling can be naturally understood in terms of hybrid YSR excitations, which are symmetric and antisymmetric linear combinations of the monomer excitations.

We corroborate this interpretation by \didv maps of the YSR wave functions taken at the corresponding peak energies in Fig.\,\ref{Fig:figure3}b. Two maps 
($\SI{2.03}{\milli\volt}$ and $\SI{1.86}{\milli\volt}$) can be assigned to the hybridized $\beta$ states. This assignment is supported by the shapes of the YSR wave functions, see close-ups in Fig.\,\ref{Fig:figure3}d. The $\beta$ state exhibits intensity at the sides and vertices of a triangular shape. Both features reappear in the patterns of the dimer, albeit with characteristic modifications due to hybridization. The map at $\SI{2.03}{\milli\volt}$ shows increased intensity in between the Fe atoms and is identified with the symmetric combination. The map at $\SI{1.86}{\milli\volt}$ exhibits a nodal line of suppressed intensity perpendicular to the dimer axis (dashed line in Fig.\,\ref{Fig:figure3}b), consistent with the antisymmetric combination. We label these dimer (d) states as $\beta^{\mathrm{dn}}$, where $\mathrm{n}$ counts the number of nodal planes. 

We also identify symmetric and antisymmetric combinations of the $\alpha$ states, which appear as a peak with a shoulder in spectra taken directly above the Fe atoms
(Fig.\,\ref{Fig:figure2}c) and can be disentangled as two distinct resonances by the corresponding modulations along the dimer axis (Fig.\,\ref{Fig:figure2}d). For this assignment of the symmetric and antisymmetric combination, we inspect the corresponding \didv maps at positive and negative bias voltage (shown in Fig.\,\ref{Fig:figure3}b and Supplementary Fig.\,7, respectively). \didv maps of $\alpha^{\mathrm{d0}}$ and $\alpha^{\mathrm{d1}}$ indeed lack or exhibit a nodal line perpendicular to the dimer axis. 

Surprisingly, we observe that the characteristic spatial shapes have switched bias polarity. Inspection of the \didv pattern probed at negative bias in the close vicinity of the Fe atoms reveals shapes which are strikingly similar to the monomer's $\alpha$ state at positive bias voltage. This is seen most clearly from the close-ups of $\alpha$ as well as $\alpha^{\mathrm{d0}*}$ and $\alpha^{\mathrm{d1}}$ shown in Fig.\,\ref{Fig:figure3}d. (Note that for $\alpha^{\mathrm{d0}}$, we present the \didv map recorded at the energy of the corresponding thermal peak as indicated by the asterisk in $\alpha^{\mathrm{d0}*}$, because the map at $\alpha^{\mathrm{d0}}$ is strongly influenced by the negative differential conductance of $\alpha^{\mathrm{d1}}$, see Supplementary Fig.\,7). This reflects that the $\alpha$ state has crossed the quantum phase transition upon dimer formation. In the monomer, the $\alpha$ state is screened and its hole-like wave function is observed at positive bias. In the dimer, the $\alpha$-derived hole-like wave functions appear at negative bias voltage and are therefore unscreened. The quantum phase transition in the $\alpha$ states from screened to unscreened is associated with an increase of the effective spin of each Fe atom from $S_\mathrm{eff}=\frac{1}{2}$ to $S_\mathrm{eff}=1$.

The crossing of the quantum phase transition between monomer and dimer requires a substantial shift of the $\alpha$ state, in addition to the hybridization splitting. Moreover, this shift is opposite to that of the $\beta$ state, which moves toward the gap edge in the dimer (see Fig.\,\ref{Fig:figure3}e for a quantitative analysis of the shift ($C$) and splitting ($D$) based on deconvolved \didv spectra shown in Supplementary Fig.\,5). Previous works \cite{Ruby2018, Ding2021} have analyzed these shifts and splittings in terms of a classical-spin model. However, neither shifts larger than the splitting nor shifts in opposite directions for different YSR states appear naturally in this model. While both shift and splitting emerge from the hybridization between the YSR excitations of the two adatoms, the splitting appears in first order in the coupling, while the shift appears only in second order \cite{Rusinov1969} (see theoretical considerations in Supplementary Note 1 for further discussion).

In contrast, a large shift can appear naturally for quantum spins. In this case, unbinding the quasiparticle bound in the $\alpha$ channel costs YSR energy, but also increases the effective spin of the Fe adatom from $S_\mathrm{eff}=\frac{1}{2}$ to $S_\mathrm{eff}=1$. This increase in the adatom spins implies that the dimer gains a larger RKKY energy. On balance, it can thus be energetically favorable for the dimer to unbind the quasiparticles, as we observe for the $\alpha$-derived states.

This is further illustrated in the schematic level diagram shown in Fig.\,\ref{Fig:figure3}f, which emphasizes the effects of RKKY coupling and hybridization by showing the states of the uncoupled dimer on the left and of the coupled dimer on the right. In the ground state of the uncoupled dimer, both adatoms are in the $(Q=3,S_\mathrm{eff}=\frac{1}{2})$ state. Exciting the $\alpha$ or $\beta$ resonance of one of the adatoms excites it into $(Q=2,S_\mathrm{eff}=1)$ states. When only one of the two adatoms is excited, there are two degenerate states of the uncoupled dimer, which are split in the coupled dimer by the YSR hybridization. Such a splitting does not exist for states in which both adatoms are in the same state, resulting in a nondegenerate configuration. In addition to the splittings, there is a downward shift of the energies of the coupled dimer due to the RKKY interaction. This shift is larger for larger $S_\mathrm{eff}$. Thus, as illustrated in Fig.\,\ref{Fig:figure3}f, the ground state of the coupled dimer can emerge from an excited state of the uncoupled dimer, consistent with our observation of a quantum phase transition of the $\alpha$ resonance from the monomer to the dimer.

The essential difference between models of classical and quantum spins is that in quantum models, the quantum phase transitions are associated with Kondo-like screening of the adatom spin. Thus, binding a quasiparticle changes the spin state of the adatom only in a quantum model (see also Supplementary Note 1). For instance, when screening a spin-$\frac{1}{2}$ impurity, it forms a singlet with the quasiparticle spin and no longer points in a preferred direction. This contrasts with classical-spin models, in which the impurity spin remains aligned along a certain direction, merely binding a quasiparticle with antiparallel spin.

\subsection{Trimer} 

Forming a trimer by deliberately adding a third atom at a distance of $3a$, we now identify three states each in the $\alpha$ and $\beta$ regions (see spectra in Fig.\,\ref{Fig:figure2}e,f). We can again disentangle the underlying hybrid states based on energy considerations, the intensity modulations of the split states along the trimer axis (see close-ups in Fig.\,\ref{Fig:figure2}g,h), and their corresponding \didv maps (Fig.\,\ref{Fig:figure3}c; in addition to the number $\mathrm{n}$ of nodal planes, we label the trimer by $\mathrm{t}$). Despite the complexity of the patterns due to the oscillatory nature of the YSR wave functions, we tentatively identify the number of nodal planes perpendicular to the trimer axis in the individual maps of the YSR resonances. In the maps at $\SI{1.57}{\milli\volt}$ and $\SI{1.84}{\milli\volt}$, we find a single nodal plane at the center of the trimer marked by a dotted line. Two nodal planes between the Fe atoms, marked by dashed lines, are clearly seen in the map at $\SI{2.02}{\milli\volt}$ and less pronounced at $\SI{1.79}{\milli\volt}$ (see also maps at negative bias in Supplementary Fig.\,7). The latter state is very close in energy to the state at $\SI{1.84}{\milli\volt}$ and may thus be partially obscured. The maps at $\SI{2.11}{\milli\volt}$ and $\SI{-1.65}{\milli\volt}$ lack clear evidence of nodal planes. We can now assign resonances to $\alpha$ and $\beta$ (see Fig.\,\ref{Fig:figure3}c) based on their energy, supported by a comparison of structural elements of their maps to monomer and dimer. In particular, the near field of $\beta^{\mathrm{t2}}$ closely resembles the ones of $\beta$ and $\beta^{\mathrm{d1}}$. We note that we assign the binding energy of the resonance $\alpha^{\mathrm{t0}}$ to have opposite sign relative to the others. This assignment is corroborated by the observation that during band formation with increasing chain length, $\alpha$-derived states appear near zero energy (see below). 

Our tentative assignments allow us to track the energy shifts and splittings from deconvolved data, which we find to be consistent with the scenario of quantum spins subject to RKKY interactions and hybridization (Fig.\,\ref{Fig:figure3}e). While the RKKY interaction leads to further shifts, hybridization increases the number of states. Importantly, the energy of the hybrid $\beta$ states is not monotonous with the number of nodal planes as we find the $\beta^{\mathrm{t2}}$ state in the energetic center of the respective triplet structure. This change in energetic order can be rationalized by accounting for strong next-nearest-neighbor RKKY interactions which arise naturally in a quantum spin model.

\subsection{Evolution of YSR bands in Fe atom chains}

\begin{figure*}[htbp]\centering
\includegraphics[width=0.8\linewidth]{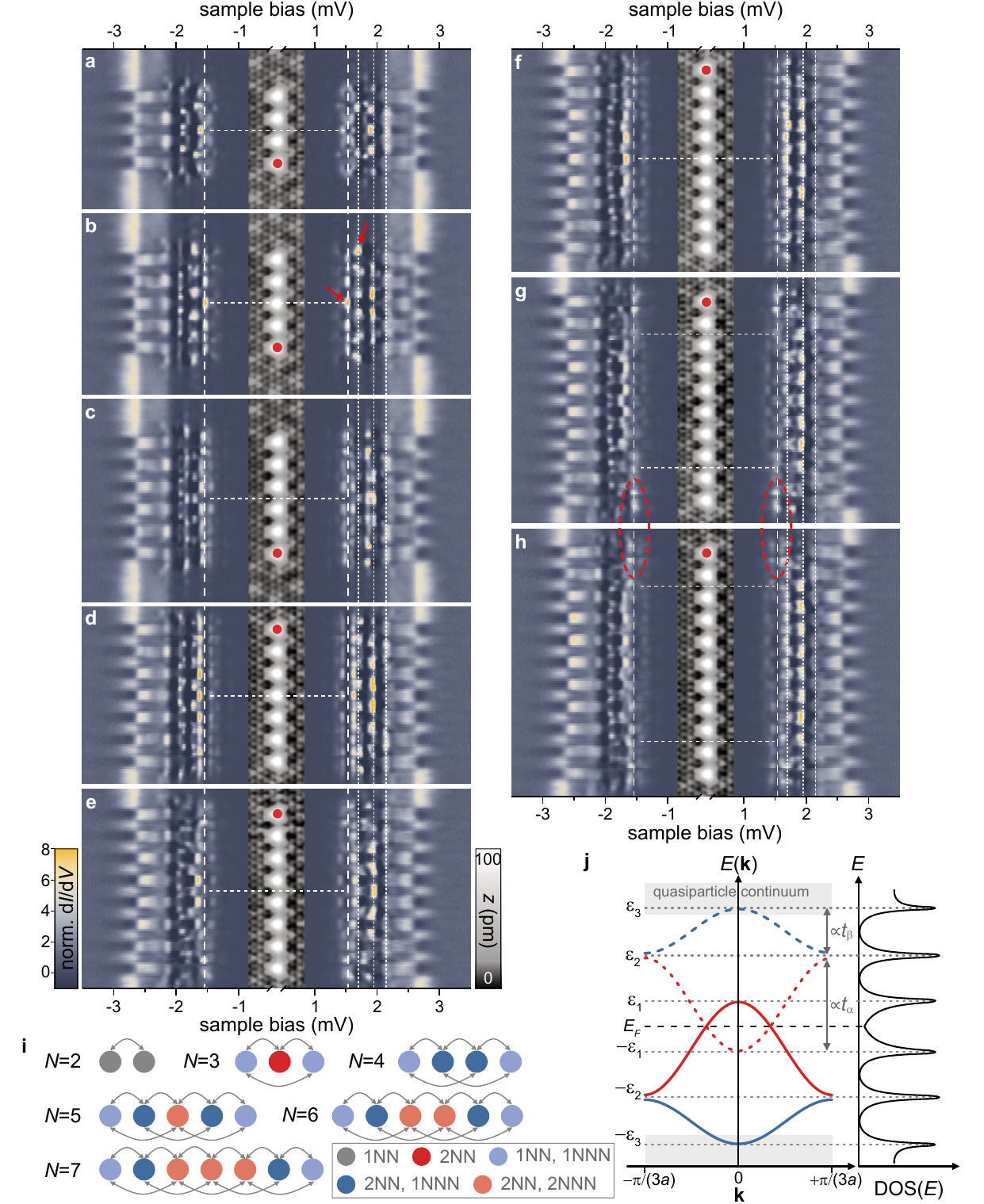}
\caption{\textbf{Evolution of YSR bands in Fe atom chains. } \textbf{a-h} Stacked constant-height \didv spectra (normalized) recorded along lines across chains containing 4-11 atoms as illustrated in the inset topographies (constant-current mode with set point $\SI{100}{\pico\ampere},\SI{10}{\milli\volt}$). Set point for spectra $250$-$\SI{700}{\pico\ampere},\SI{5}{\milli\volt}$ with a modulation of $\SI{15}{\micro\volt}$. Vertical dashed and dotted lines indicate $\Delta_{\mathrm{tip}}\approx\SI{\pm1.55}{\milli\electronvolt}$ and $\SI{2.15}{\milli\volt}$, $\SI{1.95}{\milli\volt}$, and $\SI{1.7}{\milli\volt}$ corresponding to energies of $\SI{0.6}{\milli\electronvolt}$, $\SI{0.4}{\milli\electronvolt}$, and $\SI{0.15}{\milli\electronvolt}$, respectively. These energies mark the van Hove singularities of the $\alpha$ and $\beta$ derived bands (only at positive bias for clarity), see text. The bands originating from the $\gamma$ and $\delta$ states are located within the coherence peaks and contribute to the modulations along the chain at higher biases. Horizontal dashed lines in \textbf{a-f} indicate mirror symmetry axes perpendicular to the chain.  Horizontal dashed lines in \textbf{g,h} delimit the homogeneously appearing bulk of the chain. The states at the chain's termination are marked by a red ellipse. Red dots mark the atom, that is added to the chain. Red arrows in \textbf{b} mark the large intensity at the center of the chain (left) and at the chain's termination (right), see text. \textbf{i} Schematic illustration of the number of (next) nearest (N)NN neighbors for different atoms in chains with different lengths. \textbf{j} Schematic illustration of band formation and DOS with van Hove singularities at band edges. $\pm\varepsilon_{1,2,3}$ correspond to the three pairs of dotted lines in \textbf{a-h}. 
The solid (dashed) lines represent the dispersion of the electron-like (hole-like) bands of the $\alpha$ (red) and $\beta$ (blue) bands. 
}
\label{Fig:figure4}
\end{figure*}

Figure \ref{Fig:figure4}a-h shows chains made up of $N=4$-$11$ Fe atoms (the atom added last is marked in red) and corresponding \didv spectra along their entire lengths. Neighboring atoms are placed on hollow sites with a distance of $3a$, and all adsorption sites remain close to the CDW maximum. We note that the adatoms may help to lock the CDW to the underlying atomic lattice over longer distances than for the pristine surface (cf.\ Fig.\,\ref{Fig:figure1}a and d).

In the following, we will discuss the modifications of the subgap spectra when increasing the length of the chain by the addition of individual Fe atoms. While our energy resolution no longer allows us to resolve all hybrid states derived from the individual $\alpha - \delta$ resonances, we find that the addition of single adatoms leads to changes of the states along the entire chain. Additional evidence for the delocalized nature of the subgap excitations can be obtained from the preservation of symmetries about the center of the chain as well as the convergence of spectral weight to certain energies with increasing chain length.

For the shorter chains, we observe strong modifications of the subgap spectra with the addition of every atom. These modifications become less pronounced with increasing chain length. In the presence of next-nearest-neighbor interactions, the two outermost atoms at either end of the chain are subject to different interactions from the central atoms (Fig.\,\ref{Fig:figure4}i). Thus, beginning only with $N=5$ atoms, the chains develop a uniform central region which eventually leads to the formation of band-like YSR excitations. (We do not preclude even longer-range interactions, but do not identify corresponding signatures within our resolution). 

Consistent with this picture, the subgap structure of the quadrumer (Fig.\,\ref{Fig:figure4}a) differs drastically from that of the trimer (Fig.\,\ref{Fig:figure2}f). The $\alpha$ and $\beta$ states of the monomer are now expected to form four hybrid states each. However, a detailed assignment becomes difficult as the states begin to overlap. Attaching a fifth atom causes yet another strong modification of the subgap structure (Fig.\,\ref{Fig:figure4}b). For instance, unlike for $N=4$, we now observe a delocalized state of $\alpha$ character at zero energy (i.e., at a bias of $\sim\pm\SI{1.55}{\milli\volt}$ corresponding to the tip gap). Apart from the atomic corrugation, this state has a nodeless intensity distribution exhibiting highest spectral weight at the center of the chain and falling off toward both ends (left red arrow, Fig.\,\ref{Fig:figure4}b, \didv maps are provided in Supplementary Fig.\,8). Furthermore, there are several resonances around $\pm\SI{1.7}{\milli\volt}$ which are absent for $N=4$ such as the prominent state with  strong intensity at the chain ends (right red arrow).  

Further collective changes appear upon addition of another Fe atom ($N=6$). Now, the zero-energy state is shifted to slightly higher energies. At $\sim\pm\SI{1.7}{\milli\volt}$ there is a very weak resonance with different spatial fingerprints (no intensity at the chain terminations) as opposed to $N=5$ (Fig.\,\ref{Fig:figure4}c). The concerted variation of the subgap structure along the entire chain indicates coherent coupling of all YSR excitations. This persists for the longer chains. Within the uniform center of the chain (horizontal dashed lines in panels g,h), the principal features converge into specific energy intervals as indicated by vertical dotted lines (Fig.\,\ref{Fig:figure4}). This behavior is consistent with the formation of band-like YSR excitations originating from the original $\alpha$ and $\beta$ excitations of the monomer. Once the band-like behavior is fully developed, the associated van Hove singularities become the strongest features in the \didv spectra. The interpretation of coherent band formation is further supported by the overall mirror symmetry about the center of the chain (cf.\ dashed lines in Fig.\,\ref{Fig:figure4}a-f).

The van Hove singularities allow one to extract information on the bandwidths. Each resonance of the monomer contributes an electron-like and a hole-like band of the chain. 
We suggest that the $\beta$-derived bands extend from the edge of the superconducting gap approximately halfway to the center (solid and dashed blue lines in the sketch of Fig.\,\ref{Fig:figure4}j for electron-like and hole-like bands, respectively). This assignment is based on following the uppermost and lowermost hybrid states of $\beta$-character in the \didv maps (see Supplementary Note 8 and Supplementary Figs.\,8-9). We find that the electron-like band appears at negative energies, as the $\beta$-derived states do not undergo a phase transition with increasing chain length. 

Determining the width of the $\alpha$-derived band requires tracking of its parent states for the entire sequence of chain lengths. For intermediate chain lengths, we observe $\alpha$-derived states close to zero energy (e.g. for $N=5$, see Fig.\,\ref{Fig:figure4}b). This implies that the $\alpha$-derived band crosses zero energy, so that the upper van Hove singularity of the electron-like band appears at $\sim\SI{1.7}{\milli\volt}$, i.e., at positive energies. The identification of the lower band edge is less unequivocal. We observe only three van Hove singularities (at fixed bias polarity), while four would be generically expected for two bands. Thus, the $\alpha$-derived band may either lie symmetrically around the Fermi level with band edges at $\sim\pm \SI{1.7}{\milli\volt}$ (in which case, its two van Hove singularities would be observed at the same bias voltage), or reach down to the upper van Hove singularity of the $\beta$-derived band (in which case, the second van Hove singularity of the $\alpha$ band would coincide with a van Hove singularity of the $\beta$ band). The latter scenario (red line in Fig.\,\ref{Fig:figure4}j) is more natural in view of the large observed splitting of the $\alpha$ states in the trimer (cf. Fig.\,\ref{Fig:figure3}e; note that even in the trimer, the lowest-energy $\beta$ state is already close to the highest-energy $\alpha$ state). (To demonstrate the $\alpha$-like character of the bands, we present several \didv maps in Supplementary Figs.\,8-9.)

It is interesting to note that as a result of the reduced hybridization, the bandwidths of the YSR bands in our dilute chains are considerably smaller than in densely-spaced chains \cite{NadjPerge2013}, confining bands to lie within the superconducting gap. This precludes the observation of individual states as recently done for dense chains of Mn adatoms on Nb(110), giving patterns reminiscent of particle-in-a-box states \cite{Schneider2021}.

Crossing of YSR bands through the Fermi level -- as we observe for the $\alpha$-derived band -- is a necessary requirement for realizing topological superconductivity and Majorana end states \cite{NadjPerge2013,Pientka2013}. Here, we neither find signatures of Majoranas nor of the required opening of a $p$-wave gap around the Fermi energy. The observation  of a zero-energy state localized at the chain ends for $N=10$ (Fig.\,\ref{Fig:figure4}g, red ellipses) disappears upon addition of another Fe atom (Fig.\,\ref{Fig:figure4}h) and should thus be attributed to small spectral changes at the terminations of finite chains (see Supplementary Fig.\,9 for \didv maps). 

We emphasize that the formation of bands is not a priori obvious for chains of quantum spins. Coupling of the quantum spins along with the associated subgap YSR quasiparticles in general constitutes a many-body problem \cite{Steiner2021}. 
Exciting a YSR resonance by tunneling into the quantum spin chain locally modifies the effective spin \cite{Steiner2021}. For instance, exciting the (screened) $\beta$ resonance increases the effective spin of the local adatom by $1/2$, introducing a mobile defect into the spin chain. For a ferromagnetic spin background, the mobile defect effectively leads to a band-like spectral function with van Hove singularities. As illustrated by model calculations for a spin-$\frac{1}{2}$ chain shown in Supplementary Note 2, other spin orders typically exhibit more involved excitation spectra. Our observation of YSR bands and van Hove singularities therefore suggests that the adatom spins couple ferromagnetically. (We note that strictly speaking, we cannot distinguish ferromagnetic coupling from weak spiral modulations as discussed for instance in Ref.\ \cite{Christensen2016}.) We also note that consistent with the data, our model calculations reveal strong spectral variations as a function of $N$ for short chains which eventually evolve into band-like behavior in longer chains.

 \subsection{Longer Fe chains and CDW}
 
\begin{figure*}[tb]\centering
\includegraphics[width=0.90\linewidth]{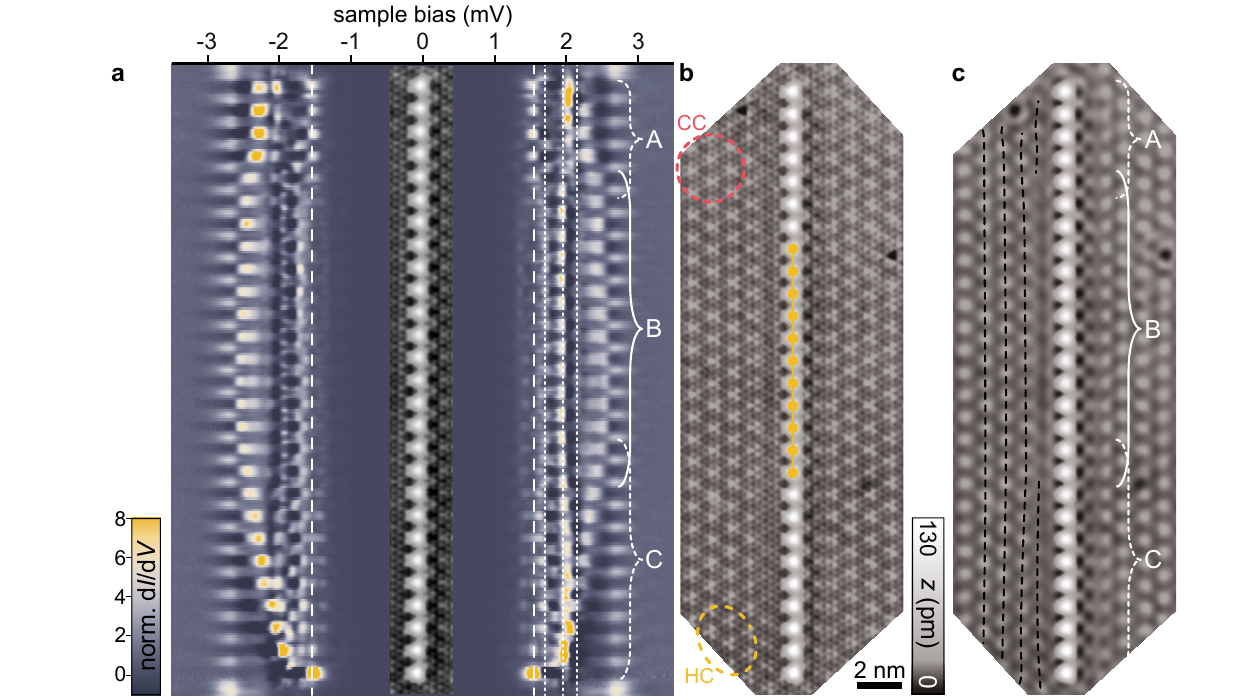}
\caption{\textbf{Band banding in longer Fe chains. }
\textbf{a} Stacked constant-height \didv spectra (normalized) recorded along a line across the 27-atom chain as illustrated in the inset topography. Set point for the spectra is $\SI{700}{\pico\ampere},\SI{5}{\milli\volt}$ with a modulation of $\SI{15}{\micro\volt}$. The (Nb) tip gap $\Delta_{\mathrm{tip}}\approx\SI{1.55}{\milli\electronvolt}$ is indicated by white dashed lines. White dotted lines are at the same energies as in Fig.\,\ref{Fig:figure4} and indicate the van Hove singularities of the $\alpha$- and $\beta$-derived bands. \textbf{b} Larger range STM topography of the 27-atom chain. The position of the former 11-atom chain (Fig.\,\ref{Fig:figure4}h) is indicated in yellow (constant-current set point is $\SI{100}{\pico\ampere},\SI{10}{\milli\volt}$). A hollow-centered (HC) region of the CDW
is highlighted by a yellow circle, a Se-centered (CC) region is marked in red. \textbf{c} A FFT-filter has been applied to the STM topography in \textbf{b} to remove the atomic corrugation. Dashed lines on the left of the chain serve as guide to the eye to illustrate the CDW order in the vicinity of the chain. Different sections of the chain are labeled with A-C. 
}
\label{Fig:figure5}
\end{figure*}

\begin{figure*}[tp]\centering
\includegraphics[width=0.90\linewidth]{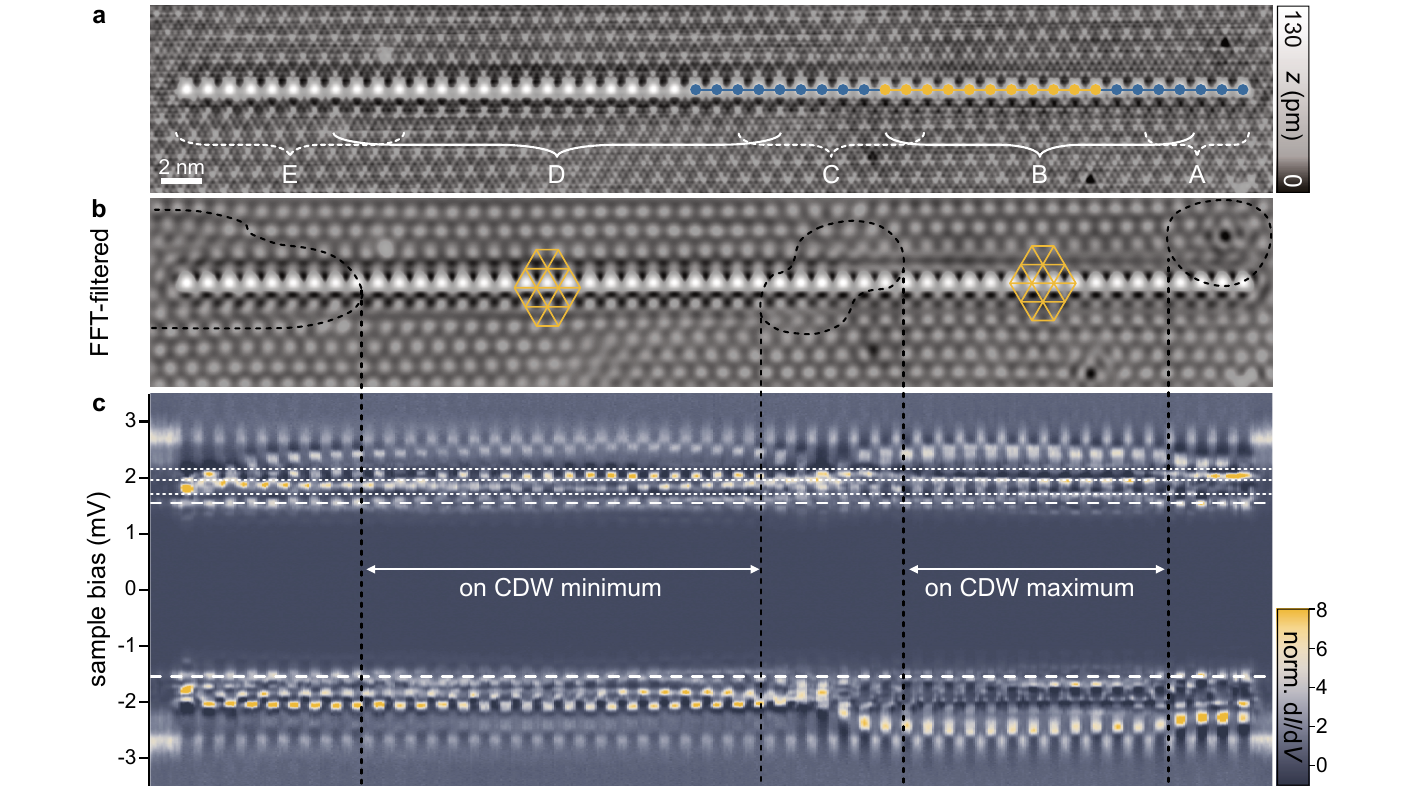}
\caption{\textbf{YSR bands across different domains of the CDW. }
\textbf{a} STM topography (constant-current mode with set point $\SI{100}{\pico\ampere},\SI{10}{\milli\volt}$) of the 51-atom chain. The position of the former 11 (27)-atom chain (Figs.\,\ref{Fig:figure4}h and \ref{Fig:figure5}) is indicated in yellow (blue). Different sections of the chain are labeled with A-E. Sections A and B are identical to Fig.\,\ref{Fig:figure5}. \textbf{b} To remove the atomic corrugation of the STM image \textbf{a} was FFT-filtered. Dashed lines encircle the most pronounced CDW distortions in the background. The yellow grids indicate the CDW and the positions of atoms within the chain relative to the CDW: crossings of the lines lie on CDW maxima, upwards pointing triangles on CDW minima. \textbf{c} Stacked constant-height \didv spectra (normalized) recorded along a line across the 51-atom chain (set point $\SI{700}{\pico\ampere},\SI{5}{\milli\volt}$ with a modulation of $\SI{15}{\micro\volt}$). The tip gap $\Delta_{\mathrm{tip}}\approx\SI{1.55}{\milli\electronvolt}$ is indicated by dashed lines. The dotted lines are located at the same energies as in Fig.\,\ref{Fig:figure4}.
}
\label{Fig:figure6}
\end{figure*}

For longer chains beyond $N=11$, the incommensurate nature of the CDW becomes increasingly relevant and the atoms can no longer all sit on maxima of the CDW. \didv spectra recorded along a 27-atom chain (Fig.\,\ref{Fig:figure5}a) reveal spatial variations of the YSR band structure. In the center of the chain (region B), the spectra are similar to those of the 11-atom chain (original 11-atom chain is indicated by yellow circles). However, the YSR bands that fall within the coherence peaks in region B (indicated in Fig.\,\ref{Fig:figure5}a) shift continuously toward the Fermi level at both ends of the chain (more pronounced in region C than in region A; for an assignment of the character of the bands, see Supplementary Figs.\,10-11). The effect of the CDW is also reflected in the loss of overall mirror symmetry about the center of the chain, which was present in shorter chains (cf.\ horizontal dashed lines in Fig.\,\ref{Fig:figure4}a-f).

The bending of the YSR bands can be correlated with changes of the CDW along the chain (Fig.\,\ref{Fig:figure5}b,c). The CDW appears as apparent height modulations  in the STM topography, which are superimposed on the atomic corrugation, see Fig.\,\ref{Fig:figure5}b. To highlight the CDW, we remove the atomic lattice from the STM image by Fourier filtering (Fig.\,\ref{Fig:figure5}c) and connect maxima of the CDW by black dashed lines (shown only to the left of the chain for clarity). In region A, the CDW is distorted by the dark defect visible at the top. In region B, the CDW runs parallel to the chain and appears locked to the atomic lattice, with all Fe atoms located on CDW maxima. In region C, the CDW does not run parallel to the chain (see additional dashed line), resulting in variations in the adsorption sites of the Fe atoms with respect to the CDW. As shown previously \cite{Liebhaber2020}, such changes in the adsorption site shift the YSR states of individual Fe atoms due to variations in the density of states and the potential scattering. Correspondingly, the CDW imposes a smoothly varying potential onto the YSR bands, which causes the YSR bands to bend. Interestingly, the YSR bands close to the superconducting gap edge appear to be more strongly affected. The bands eventually shift to zero energy at one end of the chain (region C). 
The apparent near-zero-energy state (see Supplementary Figs.\,10-11) should not be interpreted as a Majorana mode and arises from the effective band bending. 

The robustness of the band bending can be probed by further increasing the chain length. Figure\,\ref{Fig:figure6} shows a 51-atom chain, which was created by adding Fe atoms to the 27-atom chain in regions D-E. While the \didv spectra along the chain remain unchanged in regions A and B, the CDW in region C changes during manipulation (see also \didv maps in Supplementary Fig.\,12). Moreover, the \didv spectra in regions D and E differ significantly from those in region B. Here, the Fe atoms are located on CDW minima as indicated by yellow triangles in the Fourier-filtered topography in Fig.\,\ref{Fig:figure6}b, where the YSR spectrum of the monomers is characteristically different. Slow variations of the bands along the chain can again be attributed to changes in the relative alignment with respect to the CDW, with the Fourier-filtered image revealing local distortions of the CDW pattern (most pronounced distortions encircled by black dashed lines).

\section{Discussion}

Dilute chains of magnetic adatoms on superconductors are qualitatively distinct from dense adatom chains which were previously investigated in searches for topological superconductivity and Majorana bound states and in which bands of adatom $d$ orbitals are believed to cross the Fermi energy of the substrate superconductor.
In dilute chains, the quantum spins associated with magnetic adatoms induce subgap quasiparticle excitations, and the substrate-induced coupling between adatoms involves spin-spin coupling via the RKKY and Dzyaloshinsky-Moriya interactions as well as hybridization of subgap quasiparticle excitations. As a result, dilute chains provide an intriguing model system probing the correlated dynamics of the quantum spins of the adatoms and the fermionic subgap quasiparticles. 

We have realized such chains using atom manipulation with the STM tip, placing Fe atoms such that they reside on maxima of the CDW, and have probed the spatially resolved excitation spectra all along these chains. By carefully analyzing the subgap spectra exploiting both the YSR energies and YSR wave functions, we uncover evidence for the interactions between quantum spins and fermionic subgap excitations. In particular, we find that RKKY interactions are instrumental in inducing a quantum phase transition. The underlying competition between the energy of the YSR excitation and the RKKY coupling between adatom spins exists only for quantum spins and is absent for classical-spin models. Moreover, our observation of hybridization extending throughout longer chains and the ensuing formation of band-like subgap excitations suggests that the coupling between adatom spins is ferromagnetic. Our results therefore confirm that dilute chains of magnetic adatoms are not only interesting as potential platforms for the observation of Majorana bound states, but also as highly flexible realizations of a broad class of correlated spin-fermion systems. The adatom spacing as well as the orientation relative to the crystallographic axes constitute interesting parameters to be explored.
Clearly, this is not restricted to chains, but could be readily extended to two-dimensional arrangements. In both cases, it might be particularly interesting to realize antiferromagnetic RKKY coupling by judiciously adjusting the inter-adatom distance. 

The incommensurate nature of the CDW in \nbse\ does not play a significant role up to intermediate chain lengths ($\lesssim \SI{10}{\nano\meter}$) as the Fe atoms help to lock the phase of the CDW into registry with the chain. It is only for yet longer chains that the energy stored in these distortions of the CDW can no longer be sustained, triggering an abrupt change of the CDW into a different phase. Adatom chains extending across CDW domain walls exhibit significantly different band structures. In regions of smooth changes of the CDW, the YSR bands shift continuously in energy. More generally, one can exploit substrates structured by charge density waves or by moir\'e lattices to investigate domain walls between different parameter regimes of the adatom chain. Another intriguing substrate property may be a pair-density wave as putatively observed for \nbse\ \cite{Liu2021}. This might, however, require a different substrate as its effect is presumably negligible for \nbse, where it modulates the superconducting gap with the same wave vector as the CDW and by less than 1\% \cite{Liu2021}.

It would clearly be highly desirable to complement our experiments by spin-resolved STM measurements. Normal-metal-based magnetic tips have much lower resolution, making it difficult to resolve the spin polarization of the subgap resonances. A promising path forward would exploit a superconducting tip with a magnetic impurity, with its magnetization direction stabilized by a sufficiently strong magnetic field. Such a magnetic tip would be suitable for high-resolution measurements of YSR states \cite{Schneider2021}. However, care has to be taken that the magnetic field does not affect the superconductivity of the substrate, e.g., by inducing vortices or reducing the gap, thereby modifying the delicate balance between magnetic and superconducting energies of the dilute chains.


\section{Methods}
Bulk $2H$-\nbse\ crystals were grown by iodine vapor transport \cite{Rahn2012} and cleaved under ultra-high vacuum conditions. Fe atoms were deposited on the freshly cleaved sample at temperatures below $\SI{15}{\kelvin}$. Scanning tunneling microscopy and spectroscopy experiments were performed at $<\SI{1.2}{\kelvin}$. We used superconducting Nb tips with a superconducting energy gap of $\Delta_{\mathrm{tip}}\approx\SI{1.55}{\milli\electronvolt}$. These were prepared by indenting a NbTi wire into a Nb(111) substrate. The superconducting tip effectively increases the energy resolution beyond the temperature-limited Fermi-Dirac distribution. The convolution of tip and sample density of states leads to a shift of all spectral features by the tip's energy gap. To determine the precise position of the subgap states, we deconvolve the spectra and fit them by the appropriate number of Gaussian peaks, for details see Supplementary Note 6. Importantly, Nb tips are suitable for controlled atom manipulation, which allows for atom-by-atom construction of the Fe chains on the \nbse\ substrate and for tracing the spectral evolution upon chain extension. 
\\

\textbf{Note added:} After submission of our manuscript another paper investigating adatom chains in the dilute limit has been posted \cite{Kuester2021a}.

\section{Data Availability}
All data needed to evaluate the conclusions in the paper are present in the paper and the Supplementary Information.
The STM data generated in this study have been deposited in the Refubium database under accession code https://doi.org/10.17169/refubium-34026.

\section{Code Availability}
The code for the theoretical calculations shown in the Supplementary Information has been deposited in the Refubium database under accession code https://doi.org/10.17169/refubium-34026.

\def\urlprefix{}
  \def\url#1.{}

\section{Acknowledgments}
We acknowledge discussions with C.\ Mora as well as financial support by Deutsche Forschungsgemeinschaft through grant CRC 183 (project C03), by the European Research Council through the consolidator grant ``NanoSpin", and by the IMPRS ``Elementary Processes in Physical Chemistry".

\section{Supplementary Materials}
Supplementary Information accompanies this paper.

\section{Author contributions}
E.L. and L.M.R. carried out the experiments with help of G.R.. J.F.S. and F.v.O. contributed theoretical considerations and model calculations of adatom chains. S.R. and K.R. grew the \nbse\ samples. F.v.O and K.J.F. designed the project, K.J.F guided the experiments.  E.L., L.M.R., G.R., F.v.O., and K.J.F. analyzed the data. E.L., F.v.O., and K.J.F. wrote the paper with input from all coauthors. 

\section{Competing interests}
The authors declare no competing interests.

\clearpage
\setcounter{figure}{0}
\setcounter{section}{0}
\setcounter{equation}{0}
\setcounter{table}{0}
\renewcommand{\theequation}{S\arabic{equation}}
\renewcommand{\thefigure}{S\arabic{figure}}
	\renewcommand{\thetable}{S\arabic{table}}%
	\setcounter{section}{0}
	\renewcommand{\thesection}{S\arabic{section}}%

\onecolumngrid

\newcommand{\vsigma}{\mbox{\boldmath $\sigma$}}

\section*{\Large{Supplementary Material}}

\section{Supplementary Note 1: Theoretical considerations of the dimer}

The YSR states of the dimer are expected to exhibit both a shift and a splitting relative to the monomer. Our measurements show that the shift is comparable to ($\beta$) or even considerably larger ($\alpha$) in magnitude than the splitting \cite{SRusinov1969}. Within a classical-spin model, the shift is of higher order in the coupling between the adatoms and hence generically smaller than the splitting. Specifically, in a tight-binding theory of the YSR dimer assuming classical spins, the shift is controlled by \cite{SRuby2018}
\begin{equation}  
   C = \int d\mathbf{r} K(\mathbf{r}+\mathbf{d})\phi^\dagger(\mathbf{r})\phi(\mathbf{r}),
\end{equation}
while the splitting involves
\begin{equation}  
   D = \int d\mathbf{r} K(\mathbf{r})\phi^\dagger(\mathbf{r})\phi(\mathbf{r}+\mathbf{d}).
\end{equation}
Here, $\phi(\mathbf{r})$ is the two-component spinor of the monomer YSR state (centered at $\mathbf{r}=\mathbf{0}$), $\mathbf{d}$ denotes the distance vector between the adatoms constituting the dimer, and $K(\mathbf{r})$ is the exchange coupling between adatom and substrate electrons. When $|\mathbf{d}|$ is larger than the range of the exchange coupling $K(\mathbf{r})$, $C$ and $D$ are of different orders in the small overlap of the YSR wave functions of the dimer. While the shift $C \sim|\phi(\mathbf{d})|^2$ is quadratic in the overlap, the splitting $D \sim |\phi(\mathbf{d})|$ is linear.

In contrast, splitting and shifts are independent of one another within a quantum-spin model. While the splitting originates in the overlap of the YSR wave functions of the monomers, the shift involves the RKKY coupling between the adatom spins. 
The difference between the classical and quantum models lies in that only in the quantum model, the binding of a quasiparticle is associated with (Kondo-like) screening of the adatom spin. We begin by considering the case of a spin-$\frac{1}{2}$ adatom and comment on higher spins below. Consider first an adatom with a quantum spin. At weak exchange coupling between adatom spin and conduction electrons, the adatom spin remains unscreened and can assume two spin states: $\ket{\Uparrow}$ and $\ket{\Downarrow}$. At stronger exchange coupling, there is a quantum phase transition to a screened state which binds a quasiparticle. In this state, the adatom spin forms a singlet $\ket{\Uparrow\downarrow}-\ket{\Downarrow\uparrow}$ with the quasiparticle spin (spin states $\ket{\uparrow}$ and $\ket{\downarrow}$). In the unscreened state, the adatom spin can orient relative to a neighboring adatom spin and thereby gain RKKY energy. This does not happen in the screened state which is entirely isotropic in spin space. For contrast, consider the effect of quasiparticle binding on a classical spin. While a classical spin can also bind a quasiparticle, the quasiparticle spin is simply oriented opposite to the adatom spin, corresponding to states $\ket{\Uparrow\downarrow}$ and $\ket{\Downarrow\uparrow}$. In these states, the adatom spin remains free to orient relative to adjacent adatom spins, thereby benefitting energetically from the RKKY interaction. Thus, for a quantum spin, the RKKY energy of the dimer is distinctly different for screened and unscreened adatom spins. In contrast, the RKKY energy of a dimer of classical spins is independent of whether the adatom spins bind a quasiparticle or not.

This difference carries over into the energies of YSR resonances. These resonances are excitations between states with and without bound quasiparticles (energies $E_\mathrm{odd}$ and $E_\mathrm{even}$, respectively). The excitation energy equals the energy difference between these states, $E_\mathrm{YSR}=|E_\mathrm{even}-E_\mathrm{odd}|$. In a dimer of quantum spins, the states with and without bound quasiparticles have different RKKY energies and this difference in RKKY energies contributes to the YSR energy of the dimer in a manner that is independent of the hybridization. In contrast, the states with and without bound quasiparticle have the same RKKY energy for a classical spin, and the RKKY interaction cancels out from the YSR energy of a dimer of classical spins (although, of course, the RKKY interaction is nonzero also for classical adatom spins).

The same difference remains operative for higher adatom spins. For higher-spin adatoms, the adatom spin is coupled to multiple conduction-electron channels ($2S$ channels for a spin-$S$ adatom), each of which can bind a quasiparticle \cite{SOppen2021}. When increasing the exchange coupling within a particular channel, there is a quantum phase transition from a state without bound quasiparticle to a state with bound quasiparticle. 
For a classical adatom spin, binding or unbinding a quasiparticle leaves the effective impurity spin unchanged. For a quantum spin, each bound quasiparticle reduces the effective spin of the adatom by 1/2. Thus, for a quantum spin, the adatom spin can be screened to any (integer or half-integer) effective spin between zero and $S$. Exciting one of the YSR resonances changes the effective adatom spin by 1/2, leading to a corresponding change of the RKKY energy of a dimer.

We illustrate the change in RKKY coupling for a dimer of spin-$S$ adatoms, which are each coupled to a single conduction electron channel. In the unscreened state, the RKKY coupling takes the form $J\mathbf{S}_1 \cdot \mathbf{S}_2$. If one of the adatom spins, say $\mathbf{S}_1$ is screened, it effectively acts as a spin-$(S-\frac{1}{2})$ spin $\mathbf{S}_{1,\text{eff}}$, albeit with a prefactor that depends on $S$. In fact, the projection theorem yields
\begin{equation}
      \mathbf{S}_1 \to \frac{2(S+1)}{2S+1} \mathbf{S}_{1,\text{eff}}, 
\end{equation} 
so that the RKKY coupling in the presence of screening becomes $J'\mathbf{S}_{1,\text{eff}} \cdot \mathbf{S}_2$ with a modified exchange coupling constant $J' = J [2(S+1)/(2S+1)]$. For ferromagnetic coupling, for instance, this implies that the RKKY energies of the screened and unscreened states differ by
\begin{equation}
     \Delta E_\mathrm{RKKY} = \frac{S}{2S+1}J.
\end{equation} 
This difference in RKKY energies contributes directly to the energy of YSR resonances of dimers of quantum spins. Interestingly, it approaches a finite constant even in the limit of a large adatom spin.

\section{Supplementary Note 2: Theoretical considerations of the chains}

The results for monomer, dimer, and trimer suggest that both the hybridization and the RKKY coupling of neighboring adatoms in our adatom chains are comparable to or smaller than the superconducting gap. To gain intuition for interpreting the observations of longer chains (but shorter than the scale on which the CDW becomes relevant), we consider a minimal model of $S=\frac{1}{2}$ adatoms coupled to single-site superconductors \cite{SSteiner2021, SOppen2021},
\begin{align}
\label{eq:hamoriginalmodel}
    H = \sum_{j} \Bqty{
    \Delta\bqty{ c^{\dagger}_{j,\uparrow} c^{\dagger}_{j,\downarrow} + \textrm{h.c.} }  + \sum_{\sigma\sigma'}  c^{\dagger}_{j,\sigma}  \bqty{ V \delta_{\sigma\sigma'} + K  \mathbf{S}_{j} \cdot \mathbf{s}_{\sigma\sigma'} } c_{j,\sigma'}  - t \sum_{\sigma}\bqty{ c^{\dagger}_{j,\sigma}  c_{j+1,\sigma} + \textrm{h.c.} } + \mathbf{S}_{j} \cdot \mathrm{J}\cdot  \mathbf{S}_{j+1} },
\end{align}
Here, $\mathbf{S}$ denotes the $S=\frac{1}{2}$ adatom spins and $\mathrm{J}$ the spin-spin coupling, including both the RKKY interaction (symmetric part of $\mathrm{J}$) and the  Dzyaloshinsky-Moriya coupling (antisymmetric part). The exchange coupling between adatom spins and substrate electrons is taken to be isotropic and its strength is denoted by $K$. The substrate superconductor is reduced to a chain of single-site superconductors (annihilation operator $c_{j,\sigma}$ at site $j$ of the chain) with pairing strength $\Delta$. This description can be effectively thought of as projecting out the quasiparticle continuum and focusing on the subgap quasiparticles induced by the adatom spin. We also include a site energy $V$, which can be thought of as the strength of potential scattering of substrate electrons by the adatom. Finally, the hybridization of YSR states is captured by including a hopping term of strength $t$ between neighboring single-site superconductors. 

First consider the monomer \cite{SOppen2021}. Its eigenstates can be classified according to the fermion parity of the single-site superconductor. In the even-parity subspace (spanned by the empty and the doubly-occupied superconducting site), the monomer gains pairing energy $\Delta$, but no exchange coupling $K$. (Both electronic states spanning the even-fermion-parity subspace are spin singlets.) In the odd-parity subspace (spanned by the two singly-occupied states of the superconducting site), the monomer does not gain pairing energy, but benefits from the exchange coupling due to singlet formation between the impurity spin and the substrate electron. Thus, the even-fermion-parity ground state has energy $V-\sqrt{V^2+\Delta^2}$, while the odd-fermion-parity state has energy $V-\frac{3K}{4}$. The energy of the YSR excitation is then given by 
\begin{equation}
   E_\mathrm{YSR}=\sqrt{V^2+\Delta^2}-\frac{3K}{4}.
\end{equation}
For positive $E_\mathrm{YSR}$, the monomer is in the doublet of free-spin ground states
\begin{equation}
\ket{\pm} = \ket{S_{z} = \pm 1/2} \otimes (u + v c_{\downarrow}^{\dagger}c_{\uparrow}^{\dagger})\ket{ \textrm{vac}},
\end{equation}
as the adatom spin remains uncoupled to the substrate electrons. The YSR excitation excites the monomer into the screened-spin state 
\begin{equation}
  \ket{0} = ( \ket{\Uparrow\downarrow} - \ket{\Downarrow\uparrow}) /\sqrt{2},
\end{equation}
in which the impurity spin forms a singlet with the substrate-electron spin. For negative $E_\mathrm{YSR}$, the monomer is in the screened-spin ground state and the YSR excitation excites the free-spin doublet. In both cases, the energy of the YSR excitation is equal to $|E_\mathrm{YSR}|$. Here, double arrows denote the impurity-spin state, whereas single arrows denote the spin state of the substrate electron,
\begin{align}
    \ket{\sigma} = c_\sigma^{\dagger}  \ket{\textrm{vac}} = \gamma^{\dagger}_{\sigma} \pqty{u + v c_{\downarrow}^{\dagger}c_{\uparrow}^{\dagger}}\ket{ \textrm{vac}}.
    \label{eq:excitesigma}
\end{align}
The amplitudes
\begin{align}
    u = \sqrt{\frac{1}{2}\pqty{1 + \frac{V}{\sqrt{\Delta^2 + V^2}}}}\,\,\,\,\,\, , \,\,\,\,\,\, 
    v = \sqrt{\frac{1}{2}\pqty{1 - \frac{V}{\sqrt{\Delta^2 + V^2}}}}
\end{align}
denote  the conventional electron and hole amplitudes $u,v$ of BCS theory. They also enter the Bogoliubov operators
\begin{align}
      \gamma_{\sigma} = u c_{\sigma} + \sigma v  c^{\dagger}_{\bar{\sigma}}
\end{align}
of the subgap excitations.

Now consider coupled adatoms \cite{SSteiner2021}. The experimental results suggest that both the RKKY coupling and the YSR hybridization are smaller than the superconducting gap. Here, we thus consider the limit $\Delta,K,V \gg t,J,E_{\textrm{YSR}}$. We can then restrict the Hilbert space by retaining only the singlets $\ket{0_j}$ (i.e., the screened-spin states) and the doublets $\ket{\pm_j}$ (i.e., the free-spin states) at all sites $j$ of the chain. Within this restricted subspace, the spin-spin interaction $\mathrm{J}$ acts only between adjacent sites, which are both in the free-spin state. Moreover, the hybridization $t$ changes the fermion parity of the two participating sites, thereby enabling two processes. Neighboring free-spin and screened-spin sites can exchange positions with effective amplitude $\tilde{t} = t(u^2 - v^2) /2 = tV/2\sqrt{\Delta^2+V^2}$. Similarly, two adjacent free-spin sites with opposite spin can be converted into two screened sites (or vice versa) with amplitude  $\tilde{\Delta} = t uv  = t\Delta/2\sqrt{\Delta^2+V^2}$. 

We use exact diagonalization via the Lanczos scheme (with up to $300$ states in the excited state sectors) to study the local single-particle spectral function
\begin{align}
A_j(E) = \frac{\kappa}{\pi} \sum_{\sigma,\lambda} 
\bqty{ \frac{\abs{ \bra{\lambda} c^{\dagger}_{j,\sigma } \ket{\textrm{g.s.}}}^2 }
{\kappa^2 + (E-E_{\lambda} + E_{\textrm{g.s.}})^2}
+  \frac{\abs{\bra{\lambda} c^{\phantom{\dagger}}_{j,\sigma } \ket{\textrm{g.s.}}}^2}
{ \kappa^2 + ( E + E_{\lambda} - E_{\textrm{g.s.}})^2}  }.
\end{align}
The local single-particle spectral function is expected to describe tunneling experiments using superconducting tips in the limit of weak tip-substrate tunneling. We note that our numerical results are restricted in a variety of ways relative to the experimental system. They are limited to spin-$\frac{1}{2}$ impurities, do not resolve the local YSR wave functions or accommodate specifics of the substrate due to the single-site approximation for the superconductor, are restricted to zero temperature, and involve a phenomenological broadening $\kappa$. The assumption of zero temperature is restrictive since we deduce a spin-spin interaction $\mathrm{J}$ which is comparable to temperature. Consequently, one expects that excited states of the spin chain, involving for instance domain walls within a ferromagnetically ordered chain can play a significant role in the experimental chains.

\begin{figure}
    \centering
    \includegraphics[width=0.85\linewidth]{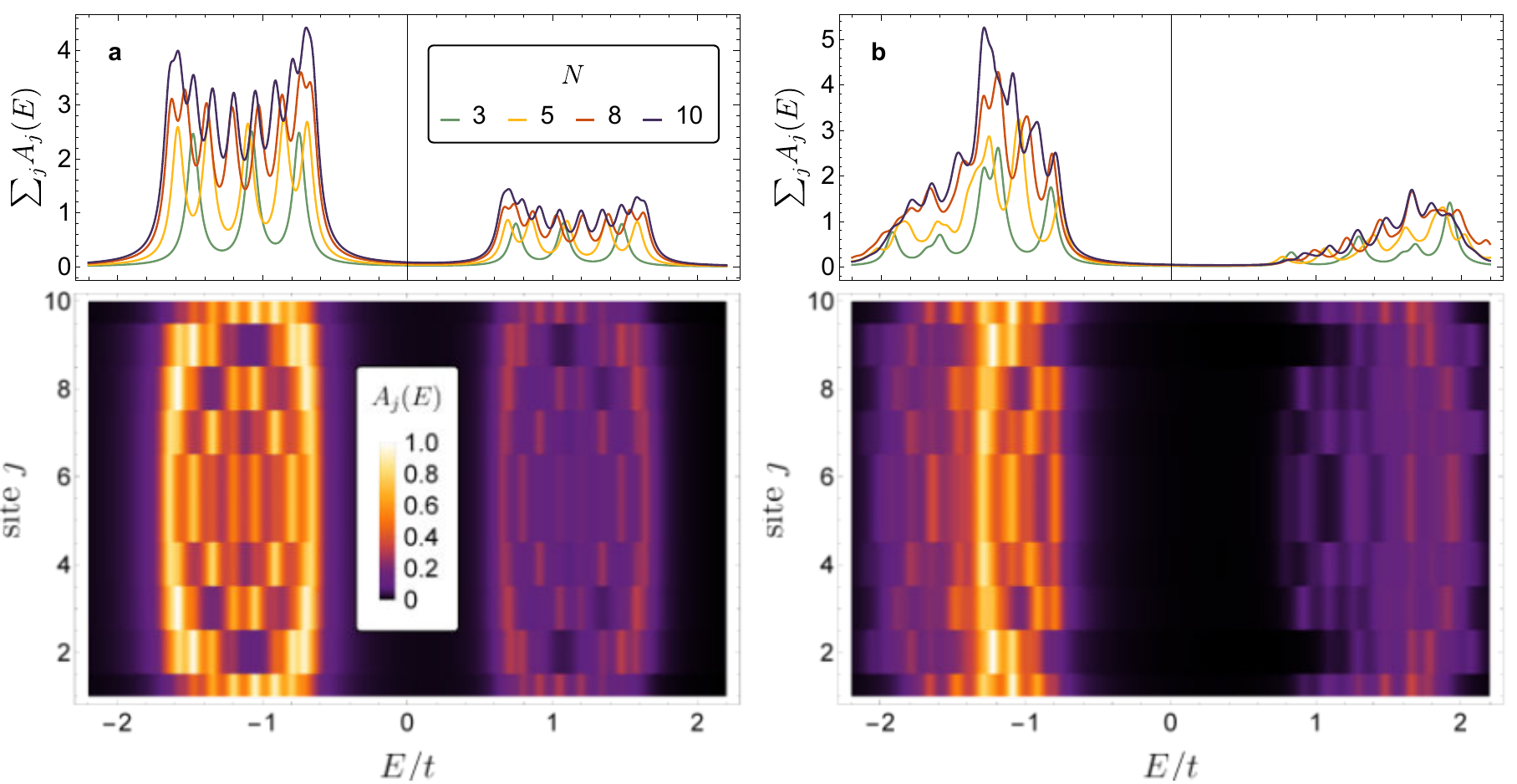}
		\caption{\textbf{Spectral functions of unscreened spin-$\frac{1}{2}$ adatom chains. }
		Local single-particle spectral density for a chain of spin-$\frac{1}{2}$ adatoms as modeled by the Hamiltonian in Supplementary Eq.\ (\ref{eq:hamoriginalmodel}). Top panels show single-particle spectral function integrated over the entire chain for different chain lengths $N$. Bottom panels exhibit spatially resolved spectral function for $N=10$. \textbf{a} Ferromagnetic phase (fully unscreened phase) with $E_{\textrm{YSR}} = t$ $J = -0.3t$ and a miniscule magnetic field $B_z = 0.001t$. \textbf{b} Antiferromagnetic phase (fully unscreened phase) with $E_{\textrm{YSR}} = t$, $J = 0.3t$. Other parameters applicable to all panels: $V = -0.6\Delta$ (chosen such that $|v|^2 \simeq 3|u|^2$, similar to  the experimental peak heights in Fig.\ 2, main text) and $\kappa = 0.05t$. }
    \label{fig:FM_AFM}
\end{figure}

Within the quantum-spin model, energy bands and van Hove singularities tend to form in particular for ferromagnetic spin ordering. In this case, the tunneling electron changes the local spin state (by screening or unscreening the local spin), creating a mobile impurity in the spin chain. In the ferromagnetic phase, this impurity can hop to neighboring sites and  effectively forms a single-particle band with van Hove singularities. Supplementary Figure \ref{fig:FM_AFM}a shows the formation of the band as the number of adatoms $N$ is increased, as seen in the local spectral function. The lower panel exhibits the local spectral function resolved all along a $N=10$ chain. Qualitatively consistent with the experimental observation, one observes that there are incipient van Hove peaks at the two band edges. Presumably, these simulations assume better energy resolution than available in experiment, which overemphasizes the bending of the van Hove singularity along the chain due to the detailed node structure of the individual wave functions. This structure may be further reduced when including thermally excited spin states. The upper panel shows how the van Hove singularity emerges with increasing $N$, again qualitatively consistent with our experimental observations. 

\begin{figure}
    \centering
\includegraphics[width=0.9\linewidth]{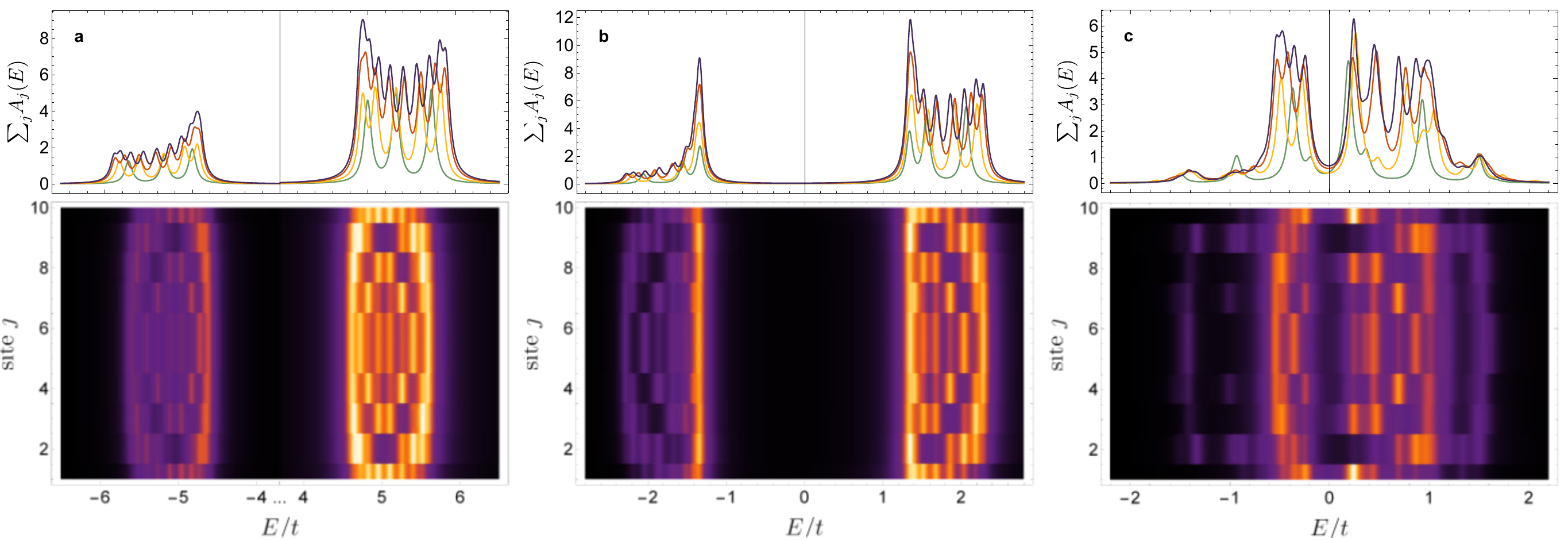}
    \caption{\textbf{Spectral functions of (partially) screened spin-$\frac{1}{2}$ adatom chains. }
		Local single-particle spectral density for a chain of spin-$\frac{1}{2}$ adatoms as modeled by the Hamiltonian in Supplementary Eq.\ (\ref{eq:hamoriginalmodel}). Top panels show single-particle spectral function integrated over the entire chain for different chain lengths $N$. Bottom panels exhibit spatially resolved spectral function for $N=10$. \textbf{a} Fully screened phase with $E_{\textrm{YSR}} = -5t$. \textbf{b} Partially screened chain with incipient singlet superconducting correlations with $E_{\textrm{YSR}} = -1.5t$. \textbf{c} Singlet superconducting phase with $E_{\textrm{YSR}} = 0$. Other parameters applicable to all panels: $V = -0.6\Delta$ (chosen such that $|v|^2 \simeq 3|u|^2$, similar to  the experimental peak heights in Fig.\ 2, main text), $J=0$, and $\kappa = 0.05t$. Color code in top panels as in Supplementary Fig.\ \ref{fig:FM_AFM}.}
    \label{fig:singletphases}
\end{figure}

For contrast, Supplementary Fig.\ \ref{fig:FM_AFM}b shows corresponding results for antiferromagnetic RKKY coupling. Unlike for ferromagnetic spin ordering, the excitation spectrum is not single-particle-like and does not exhibit pairs of van Hove singularities for any chain length. Single-particle bands and the emergence of van Hove singularities with increasing $N$ are also observed when all spins are screened, see Supplementary Fig.\ \ref{fig:singletphases}a. However, the experimental results indicate that the current system does not realize this limit (unscreened $\delta$ resonance as well as partially screened $\alpha$ band). Panels b and c of Supplementary Fig.\ \ref{fig:singletphases} show corresponding results for monomer YSR energies such that there is only partial screening. In this case, the single-particle spectral function exhibits a pair of peaks which are located at symmetric energies relative to the Fermi energy. These peaks are BCS coherence peaks induced by the singlet superconducting correlations $\tilde\Delta$. For unscreened $\delta$ resonance, this phase should not be realized in the experimental system, since ferromagnetic order suppresses the underlying singlet correlations. We also do not observe the well-developed gap around the Fermi energy which exists in this phase. 

As discussed in the main text, we assume that the $\delta$ resonance is likely unscreened. However, this assignment (based on analyzing shifts of the YSR energy with adsorption position relative to the CDW) is less definitive than the ones for $\alpha$-$\gamma$. We briefly comment on this issue in light of the observed band formation. When all YSR states are in the screened state in the monomer, the adatom spin is effectively fully screened in the monomer. Due to the quantum phase transition for the $\alpha$ resonance, we would then conclude that the adatom chains can effectively be viewed as partially filled spin-$\frac{1}{2}$ chain. In this case, there is a substantial region in the phase diagram in which the system is in a spin-singlet superconducting phase (in addition to the ferromagnetic phase discussed above) \cite{SSteiner2021}. As seen in panels b and c of Supplementary Fig.\ \ref{fig:singletphases},
the coherence peaks of this singlet superconductor naturally provide a single dominant peak which might be consistent with our observations of the $\alpha$ band. Indeed, as discussed in the main text, it is possible (though less likely) that this resonance leads to symmetric peaks about the Fermi energy. However, an interpretation of this observation in terms of coherence peaks of a spin-singlet superconductor seems unlikely since we do not observe the associated gap around the Fermi energy. Note in particular that according to the top panels, the zero-temperature spectral function does not exhibit any structure within this gap for any chain length $N$. Moreover, the observation of two peaks for the $\beta$ band is much less natural in such a spin-singlet superconductor, as its excitation spectrum does not exhibit single-particle character. Based on these considerations, we conclude that the observations on adatom chains support our identification of the $\delta$ resonance as unscreened.

\section*{Supplementary Note 3: Adsorption sites and incommensurate CDW}
\begin{figure}[ht]\centering
\includegraphics[width=0.95\linewidth]{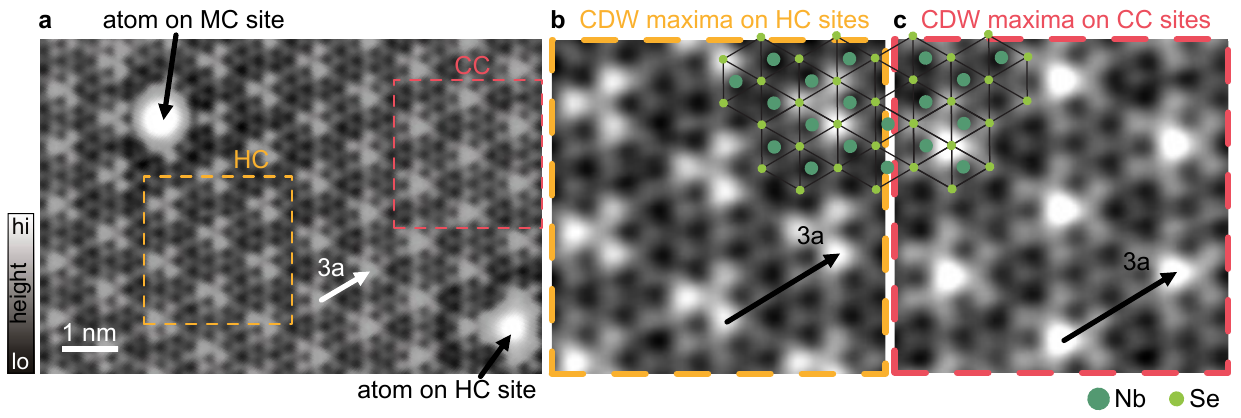}
\caption{\textbf{Adsorption sites on the incommensurate CDW.}
\textbf{a} Atomic resolution topography (constant-current mode, set point: \spat). \textbf{b} and \textbf{c} are zooms into the areas marked in \textbf{a}. Arrows indicate one CDW period ($\approx 3a$). The overlaid grid shows the terminating Se-layer (bright green) and the Nb layer beneath (dark green).
}
\label{Fig:CDW}
\end{figure}

Fe atoms deposited at low temperature on the clean \nbse\ surface adsorb in two distinct sites. They can be distinguished by their different apparent height (Supplementary Fig.\,\ref{Fig:CDW}a) and identified as sitting in the two different hollow sites of the terminating Se layer, which differ by the presence (metal-centered, MC) or absence (hollow-centered, HC) of a Nb atom underneath. The YSR states from the Fe atoms in the different adsorption sites differ substantially \cite{SLiebhaber2020}. Here, we investigate and build up chains from Fe atoms on the HC sites only. 

The incommensurate nature of the CDW with a periodicity of $\gtrsim 3a\times 3a$ (see white arrow in Supplementary Fig.\,\ref{Fig:CDW}a) is reflected in an additional modulation to the atomic corrugation in the STM images. When the maximum of the CDW is located on a hollow site of the Se layer (hollow-centered, HC), the topographic pattern appears with a three-petaled shape (yellow area in Supplementary Fig.\,\ref{Fig:CDW}a and Supplementary Fig.\,\ref{Fig:CDW}b). In contrast, when the CDW maximum lies on top of a Se atom (chalcogen-centered, CC), the STM image is petalless (red area in Supplementary Fig.\,\ref{Fig:CDW}a and Supplementary Fig.\,\ref{Fig:CDW}c). Due to the incommensurability of the CDW, the patterns smoothly transform into one another. 

\section*{Supplementary Note 4: Kondo effect}
\begin{figure}[ht]\centering
\includegraphics[width=0.95\linewidth]{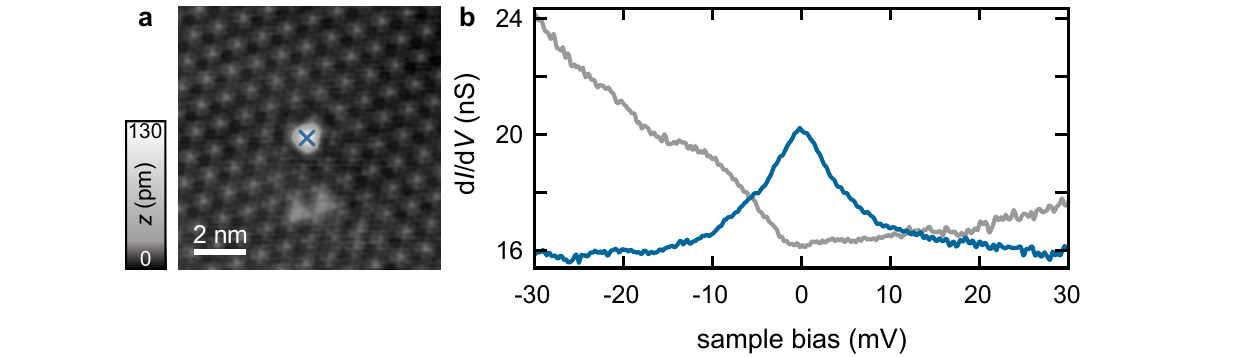}
\caption{\textbf{Kondo effect on single Fe adatom. }
\textbf{a} Topography showing a single Fe atom (constant-current mode, set point: $\SI{200}{\pico\ampere},\SI{4}{\milli\volt}$) recorded with a Pb tip. \textbf{b} Differential conductance spectra recorded on the atom (blue) and on the bare \nbse\ (gray) at $T=\SI{8}{\kelvin}$. Feedback was opened at $\SI{500}{\pico\ampere},\SI{30}{\milli\volt}$ and a modulation of $\SI{0.5}{\milli\volt}$ was used.}
\label{Fig:Kondo}
\end{figure}

In the main text, we show that the splitting and shift of YSR states of the Fe adatoms due to RKKY interactions cannot be explained with a classical-spin model, but instead requires a quantum mechanical description. Another expression of the quantum nature of the Fe adatom's spin is found when inspecting the spectroscopic signature in the normal state of the substrate. Supplementary Figure \,\ref{Fig:Kondo} shows a zero-bias peak measured on an Fe atom sitting on a hollow site of the Se lattice at a temperature of \SI{8}{\kelvin}. This resonance reveals a Kondo resonance as a fingerprint of a quantum spin on the surface. 

\section*{Supplementary Note 5: Atom manipulation using a superconducting Nb tip}
All measurements were performed with a superconducting Nb tip fabricated by indenting a NbTi-tip into a superconducting Nb sample until a sharp stable tip apex exhibiting the full Nb gap of $\Delta\approx\SI{1.55}{\milli\electronvolt}$ is achieved. To construct the adatom chains, controlled manipulation of the Fe atoms is required. The atoms could be positioned in a very precise manner after placing the STM tip in their close vicinity and dragging them across the surface, while applying small bias voltages ($\sim$mV) and currents in the $\SI{}{\nano\ampere}$-regime (exact values depend on the tip apex).

\section*{Supplementary Note 6: Deconvolution and fitting procedure}

Due to the superconducting properties of the STM tip, the \didv spectra are a convolution of the density of states of tip and substrate. To extract the density of states of the substrate, we numerically deconvolve the spectra as described in the supplementary information to Ref. \cite{SLiebhaber2020}. Deconvolved spectra of the monomer, dimer and trimer (original spectra in Fig.\,2a,c,e of the main manuscript) are shown in Supplementary Fig.\,\ref{Fig:fit}a.
To determine the shift and split of the $\alpha$- and $\beta$-derived YSR states in the dimer and trimer structure, we symmetrized the deconvolved spectra with respect to zero energy and then fitted them with the appropriate number of Gaussian peaks using the following equation: 
\begin{equation}
\mathrm{DOS}(E) =D_0 + \sum_{i=1}^N \left( A_i \, e^{-(E \pm E_{\alpha_i})^2/(2\sigma^2)}+B_i \, e^{-(E \pm E_{\beta_i})^2/(2\sigma^2)} \right). 
\label{eq:fit}
\end{equation}

Here, $D_0$ is an intensity offset, $A_i$ ($B_i$) are the (symmetric) amplitudes of the $\alpha$- ($\beta$-)derived resonances and $\sigma$ is the width of the Gaussian peaks. The number of split YSR states depends on the number of atoms N in the hybridized structure, being two states in the dimer ($\alpha^{\mathrm{d0,d1}}$ and $\beta^{\mathrm{d0,d1}}$) and three YSR states in the trimer ($\alpha^{\mathrm{t0,t1,t2}}$ and $\beta^{\mathrm{t0,t1,t2}}$). 
As shown in the main manuscript, the YSR states exhibit strong spatial intensity variations, with some peaks being hardly visible in some spectra. The most reliable determination of all peak positions thus follows from a detailed analysis of a set of spectra from the line profiles in Fig.\,2b,d,f of the main text.  Several spectra were fitted and the fit results of the YSR energies were averaged. The error bars of the YSR energies were determined from the standard deviation of the fit, the error margin of the energy gap of the tip, the modulation voltage of the lock-in, and the sampling interval of the data used for the deconvolution.

Figure\,3e (main part) compiles the resulting peak positions in the monomer, dimer and trimer. Supplementary Table\,\ref{tab:D_C} summarizes the energy splittings $D_{\alpha ,\beta}$ and shifts $C_{\alpha ,\beta}$ in the hybridized structures relative to the monomer.

\begin{figure}[ht]\centering
\includegraphics[width=0.9\linewidth]{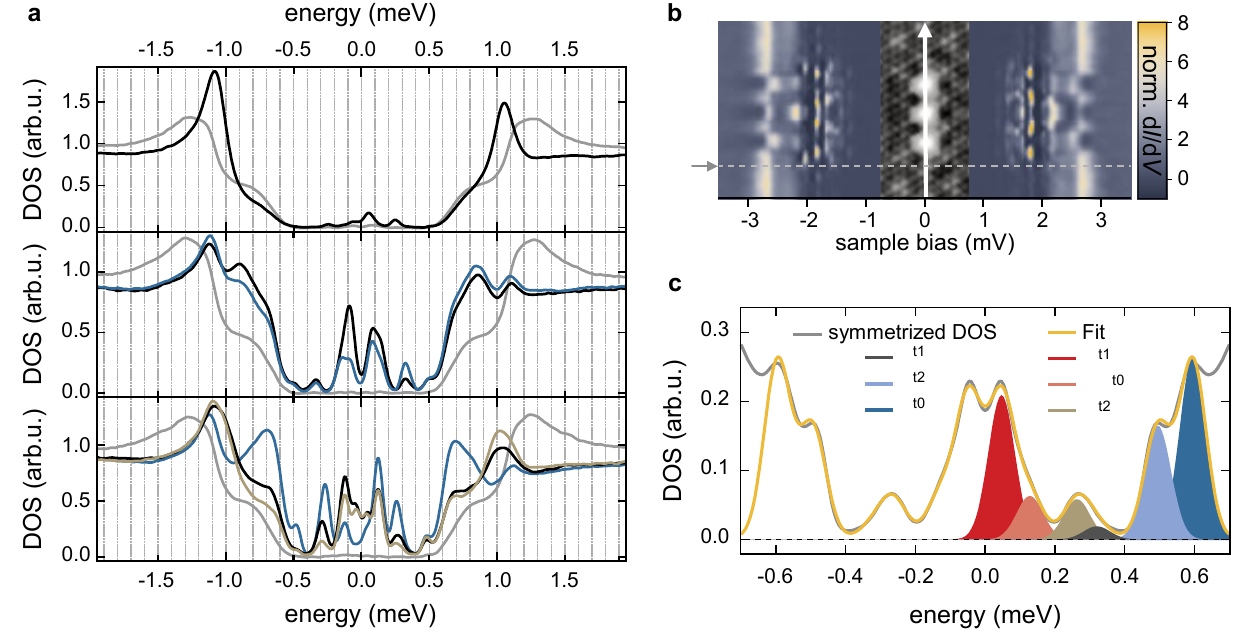}
\caption{\textbf{Illustration of deconvolution procedure. }
\textbf{a} Deconvolved data of Fig.\,2a,c,e. For the deconvolution $\Delta_{\mathrm{tip}}=\SI{1.55}{\milli\electronvolt}$ and a depairing factor of $\Gamma=\SI{5}{\micro\electronvolt}$ were used. \textbf{b} Data reproduced from Fig.\,2f. The gray dashed line marks the position of the spectrum that is deconvolved and symmetrized in \textbf{c} (gray line). The fit (yellow) and the individual Gaussian peaks (only at positive energy) are shown in red and blue colors for the $\alpha$- and $\beta$-derived states, respectively.
}
\label{Fig:fit}
\end{figure}

An example of a fitted trace of a trimer is shown in Supplementary Fig.\,\ref{Fig:fit}c (position of the original spectrum marked by the arrow and dashed line in Supplementary Fig.\,\ref{Fig:fit}b). The gray trace is the symmetrized density of states. The fit according to Supplementary Eq.\,\ref{eq:fit} is shown in yellow and the individual Gaussian peaks are shown (for positive energies only) in color. 
The fit yields reliable values for $\alpha^{\mathrm{t0}},\alpha^{\mathrm{t1}}$ and $\beta^{\mathrm{t2}}$, as they are well separated peaks. In contrast, as discussed in the main text, $\alpha^{\mathrm{t2}}$ and $\beta^{\mathrm{t1}}$ strongly overlap leading to one broad resonance (full width half maximum of $\approx \SI{100}{\micro\electronvolt}$). The peak corresponding to $\beta^{\mathrm{t0}}$ is very close to the quasiparticle coherence peaks. For these reasons, we added additional error margins of $\pm\SI{50}{\micro\electronvolt}$ to the latter three resonances.

\begin{table}
\begin{center}
\begin{tabular}{ ccccccc } 
& $|D_{\alpha} (\SI{}{\micro\electronvolt})|$ & $|D_{\beta} (\SI{}{\micro\electronvolt})|$  & $C_{\alpha} (\SI{}{\micro\electronvolt})$ & $C_{\beta} (\SI{}{\micro\electronvolt})$    \\
\hline
dimer & $69\pm 33$  & $173\pm 33$  & $+184\pm 29$ & $-160\pm 29$   \\
trimer & $388\pm 62$ & $286\pm 79$  & $+14\pm 33$ & $-201\pm 36$    \\ 
\hline
\end{tabular}
\vspace{2mm}
\caption{\textbf{Split and shift of YSR states. } Results for the splits $D_{\alpha ,\beta}$ and shifts $C_{\alpha ,\beta}$ (relative to the monomer) of the hybrid $\alpha$- and $\beta$-states of the dimer and trimer obtained from Fig.\,3e main part.
}
\label{tab:D_C}
\end{center}
\end{table}

\section*{Supplementary Note 7: Influence of the Fe atoms to the CDW}

\begin{figure}[ht]\centering
\includegraphics[width=0.95\linewidth]{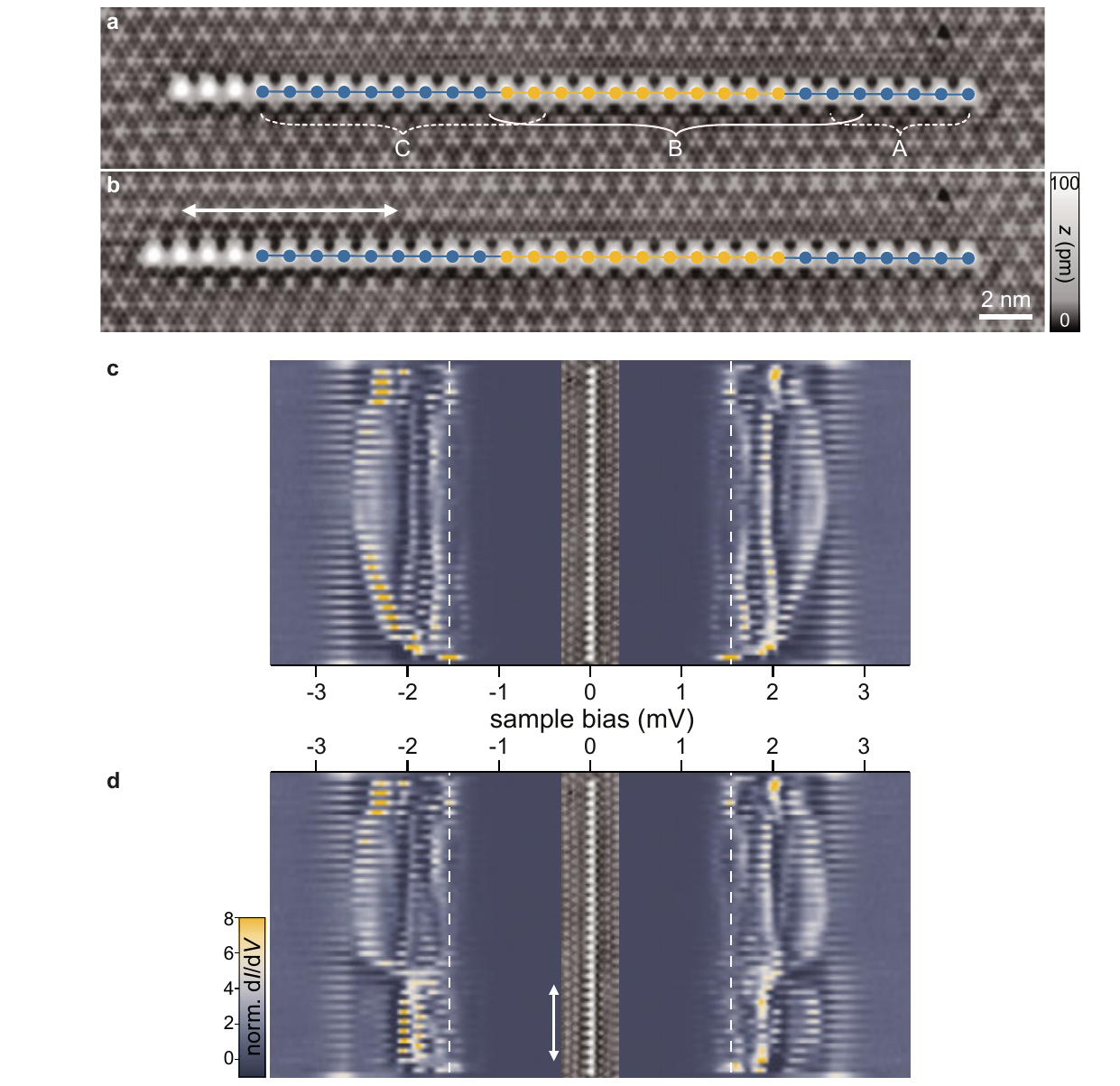}
\caption{\textbf{Switching of CDW and YSR bands across different domains. }
\textbf{a,b} Constant-current topography of the 30-atom chain \textbf{a} and 31-atom chain \textbf{b} recorded at a set point of \sptop. The former 11-atom (27-atom) chains are indicated in yellow (blue). The white arrow marks the section of the 31-atom chain where atoms are located on a CDW minimum. \textbf{c,d} Line profiles of normalized \didv spectra recorded along the 30-atom chain \textbf{c} and 31-atom chain \textbf{d} (constant-height mode, feedback opened at $\SI{700}{\pico\ampere},\SI{5}{\milli\volt}$ with a modulation of $\SI{15}{\micro\volt}$). The arrow marks the same section as in \textbf{b}. Dashed lines indicate the tip gap. 
}
\label{Fig:30-31atoms}
\end{figure}

As described in the main manuscript, an 11-atom chain can be built in such a way that all Fe atoms sit on maxima of the CDW. 
This contrasts the smooth variation of the CDW and indicates that the Fe atoms may help to lock the phase of the CDW. 
Upon further extension of the chain, the locking is not operable anymore and the atoms cannot sit on the CDW maxima. From this length on, the smoothly varying phase along the atomic lattice leads to shifts of the YSR band structure. Importantly, during the manipulation and extension of the chain up to a chain length of $N=30$ atoms, the CDW did not change abruptly. In contrast, upon attachment of the 31$^{\mathrm{st}}$ atom, an abrupt change occurred in the CDW (compare Supplementary Fig.\,\ref{Fig:30-31atoms}a and b).
This change is most clearly expressed in a change of the \didv spectra. While the 30-atom chain exhibits the smooth variations of van Hove singularities toward the chain's terminations, the 31-atom chain shows two distinct areas of YSR bands. Close inspection of the CDW phase reveals that the atoms in the area marked by a white arrow are now located on minima of the CDW (Supplementary Fig.\,\ref{Fig:30-31atoms}b,d).
These observations suggest that the Fe atoms favor adsorption sites on the maxima or minima of the CDW and thus push the CDW into the respective phase. However, when too much energy is stored in the locked CDW, stress is released by an abrupt switching of the CDW.

\section*{Supplementary Note 8: Additional \didv maps}
In the main manuscript, we have presented selected \didv maps of monomers, dimer, trimers and an extended 27-atom chain, which helped identifying the nature of the YSR states. This section provides additional data, which corroborate the assignments in the main text. 

\begin{figure}[ht]\centering
\includegraphics[width=0.95\linewidth]{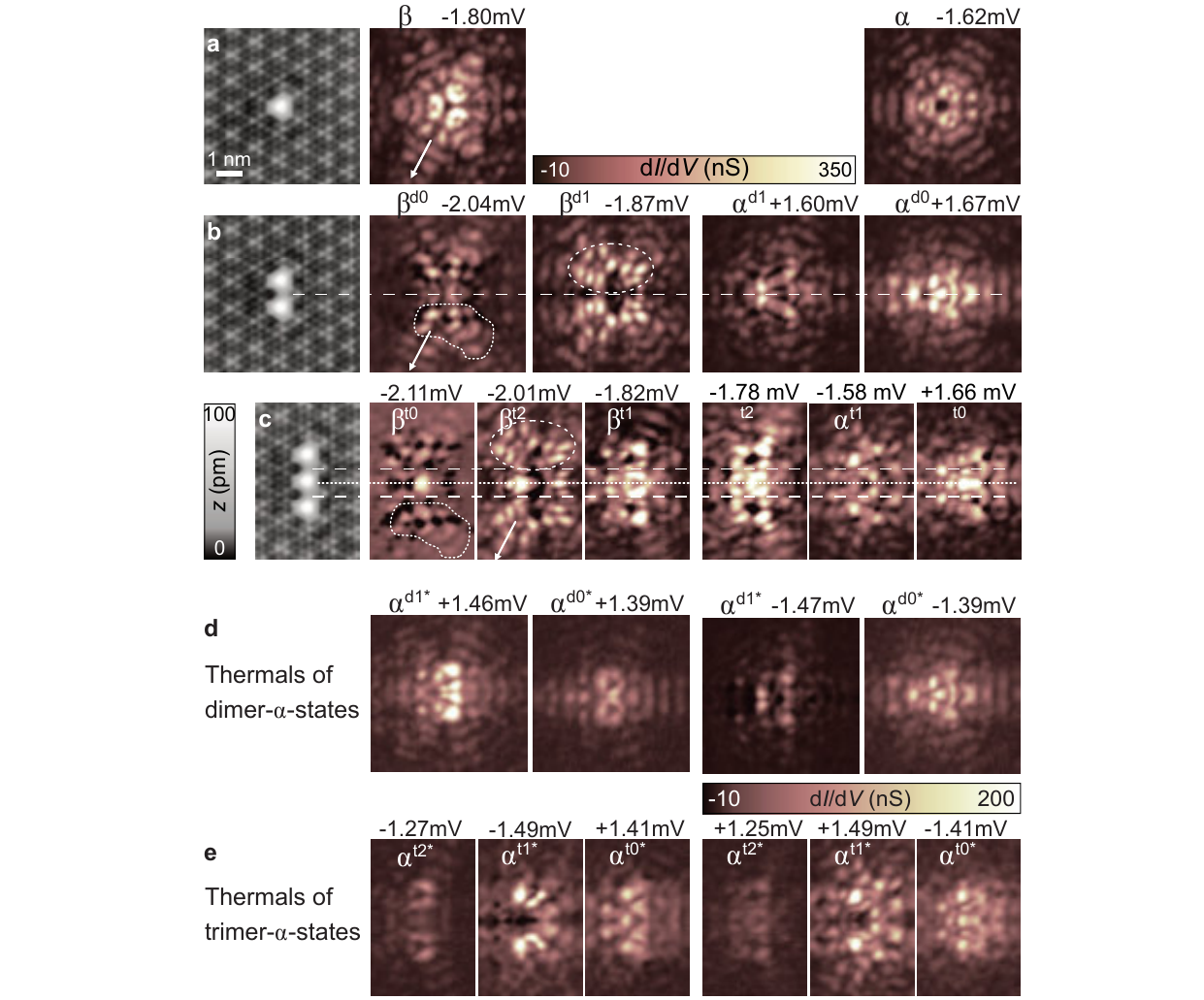}
\caption{\textbf{YSR wave functions of monomer, dimer and trimer. }
Complementary data of Fig.\,3 in the main part. \textbf{a-c} STM topographies (constant-current mode with set point \sptop) of one to three Fe atoms with spacing of $3a$ in the left. Corresponding constant-contour \didv maps of the (hybridized) YSR states in the monomer (top row), dimer (middle row) and trimer (bottom row). Bias voltages are given above each panel ($V_{\mathrm{YSR}}=\pm|\Delta_{\mathrm{tip}}+E_{\mathrm{YSR}}|/e$ with $\Delta_{\mathrm{tip}}\approx\SI{1.55}{\milli\electronvolt}$). Constant-contour feedback set point is $\SI{250}{\pico\ampere},\SI{5}{\milli\volt}$ and the modulation is $\SI{15}{\micro\volt}$. \textbf{d,e} Constant-contour maps (same parameters as in \textbf{a-c}) of the thermally activated $\alpha$-states of the dimer \textbf{d} and the trimer \textbf{e}.
}
\label{Fig:didv}
\end{figure}

Supplementary Figure \ref{Fig:didv}a-c complement the \didv maps presented in Fig.\,3 of the main manuscript with the opposite bias polarity. Additionally, Supplementary Fig.\,\ref{Fig:didv}d,e show \didv maps of the thermally activated $\alpha$-derived resonances at $eV_{\mathrm{YSR}}=\mp|\Delta_{\mathrm{tip}}-E_{\mathrm{YSR}}|$. These data support the identification of the $\alpha$- and $\beta$-derived states by their distinct symmetries and presence/absence of nodal planes. In particular, the assignment of both $\alpha$ states of the dimer to lie at the other bias polarity becomes clear when inspecting the thermal maps. The patterns of the monomer, dimer, and trimer share several features, which help in the identification of the origin of the $\beta$-derived states. For instance, oscillations along the white arrows in Supplementary Fig.\,\ref{Fig:didv}a are the same along the arrow in Supplementary Fig.\,\ref{Fig:didv}b, and thus indicate the $\beta$ nature of the dimer state at $\SI{-2.05}{\milli\volt}$. The scattering patterns encircled by white dashed lines in the dimer of Supplementary Fig.\,\ref{Fig:didv}b, can be found in the trimer again in Supplementary Fig.\,\ref{Fig:didv}c.

\clearpage

\begin{figure}[ht]\centering
\includegraphics[width=0.95\linewidth]{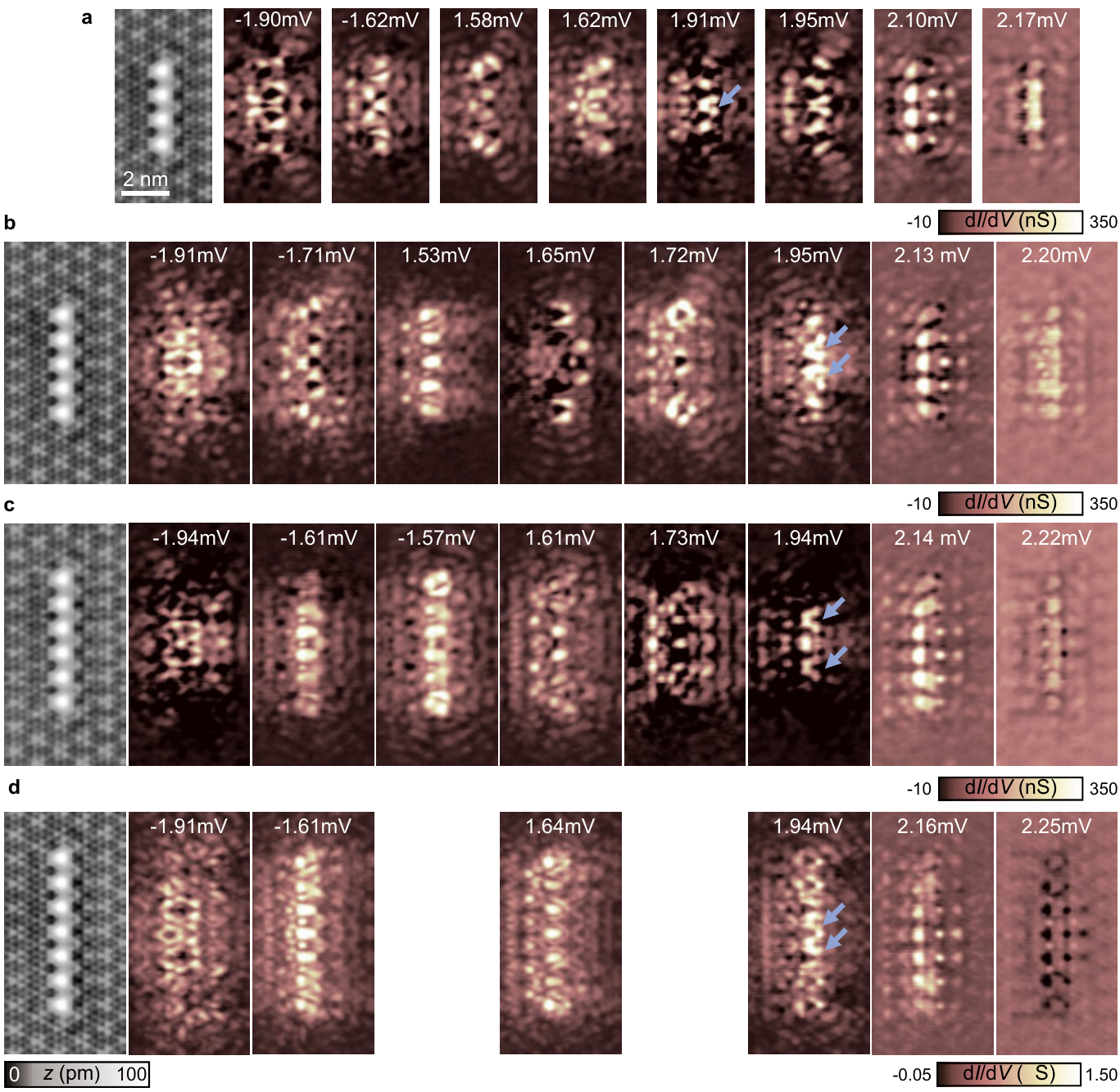}
\caption{\textbf{YSR wave functions in chains of $N=4 - 7$ Fe atoms. }
Complementary data to Fig.\,4a-d of the main manuscript. \textbf{a-d} STM topographies (constant-current mode with set point $\SI{100}{\pico\ampere},\SI{10}{\milli\volt}$) of the 4-, 5-, 6- and 7-atom chains with spacing of $3a$ on the left. Corresponding constant-contour \didv maps at selected bias voltages (values indicated in the panels) on the right. Constant-contour feedback was opened at $\SI{250}{\pico\ampere},\SI{5}{\milli\volt}$ (\textbf{a-c}) or $\SI{700}{\pico\ampere},\SI{5}{\milli\volt}$ (\textbf{d}) and a modulation of $\SI{15}{\micro\volt}$ was used ($\Delta_{\mathrm{tip}}\approx\SI{1.55}{\milli\electronvolt}$). 
}
\label{Fig:4-7atoms}
\end{figure}

\begin{figure}[ht]\centering
\includegraphics[width=0.95\linewidth]{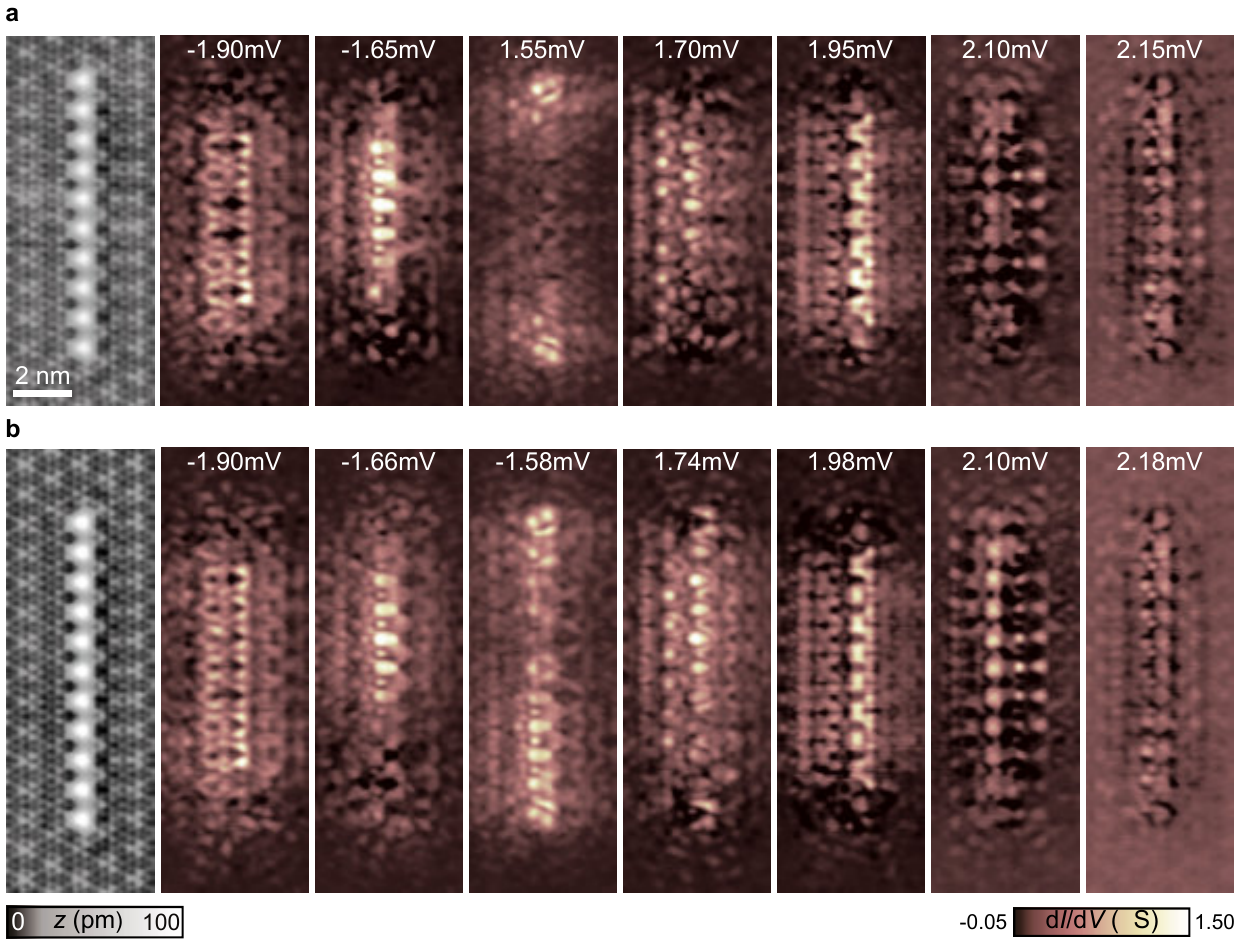}
\caption{\textbf{YSR wave functions in chains of $N=10$ and $N=11$ Fe atoms. }
Complementary data to Fig.\,4g,h of the main manuscript. \textbf{a,b} STM topographies (constant-current mode with set point $\SI{100}{\pico\ampere},\SI{10}{\milli\volt}$) of the 10- and 11-atom chains with spacing of $3a$ in the left. Corresponding constant-contour \didv maps at selected bias voltages (values indicated in the panels) on the right. Constant-contour feedback was opened at $\SI{700}{\pico\ampere},\SI{5}{\milli\volt}$ and a modulation of $\SI{15}{\micro\volt}$ was used ($\Delta_{\mathrm{tip}}\approx\SI{1.55}{\milli\electronvolt}$). 
}
\label{Fig:10-11atoms}
\end{figure}

Supplementary Figures\,\ref{Fig:4-7atoms} and \ref{Fig:10-11atoms} show additional \didv maps at various energies for chain lengths between 4 and 11 atoms. The maps reflect the character of the hybrid wave functions and their intensity distributions along shorter chains and further highlight the extended nature of the YSR bands for increasing chain length ($N>5$). 
We can identify the origin of the bands at larger chain length by structural elements common to the single atom, short and long chains. For example, we find signatures of the $\beta$-derived state within the coherence peaks always around $\gtrsim\SI{2.15}{\milli\volt}$ and around $\SI{1.95}{\milli\volt}$ as indicated by the blue arrows. States deep within the gap arise from $\alpha$-states as suggested by comparing the shapes to the $\alpha$-trimer states (Figs.\,3c main part and Supplementary Fig.\,\ref{Fig:didv}).

\clearpage

As mentioned in the main text there is a zero-energy end state ($\Delta_{\mathrm{tip}}\approx\SI{1.55}{\milli\volt}$) in the 10-atom chain. However, as can be inferred from the line spectra in Fig.\,4g,h in the main paper, this state is shifted to slightly higher energies at the 11-atom chain ($\SI{1.58}{\milli\electronvolt}$). \didv maps shown in Supplementary Fig.\,\ref{Fig:10-11atoms} sustain this interpretation.

\clearpage

\begin{figure}[ht]\centering
\includegraphics[width=0.95\linewidth]{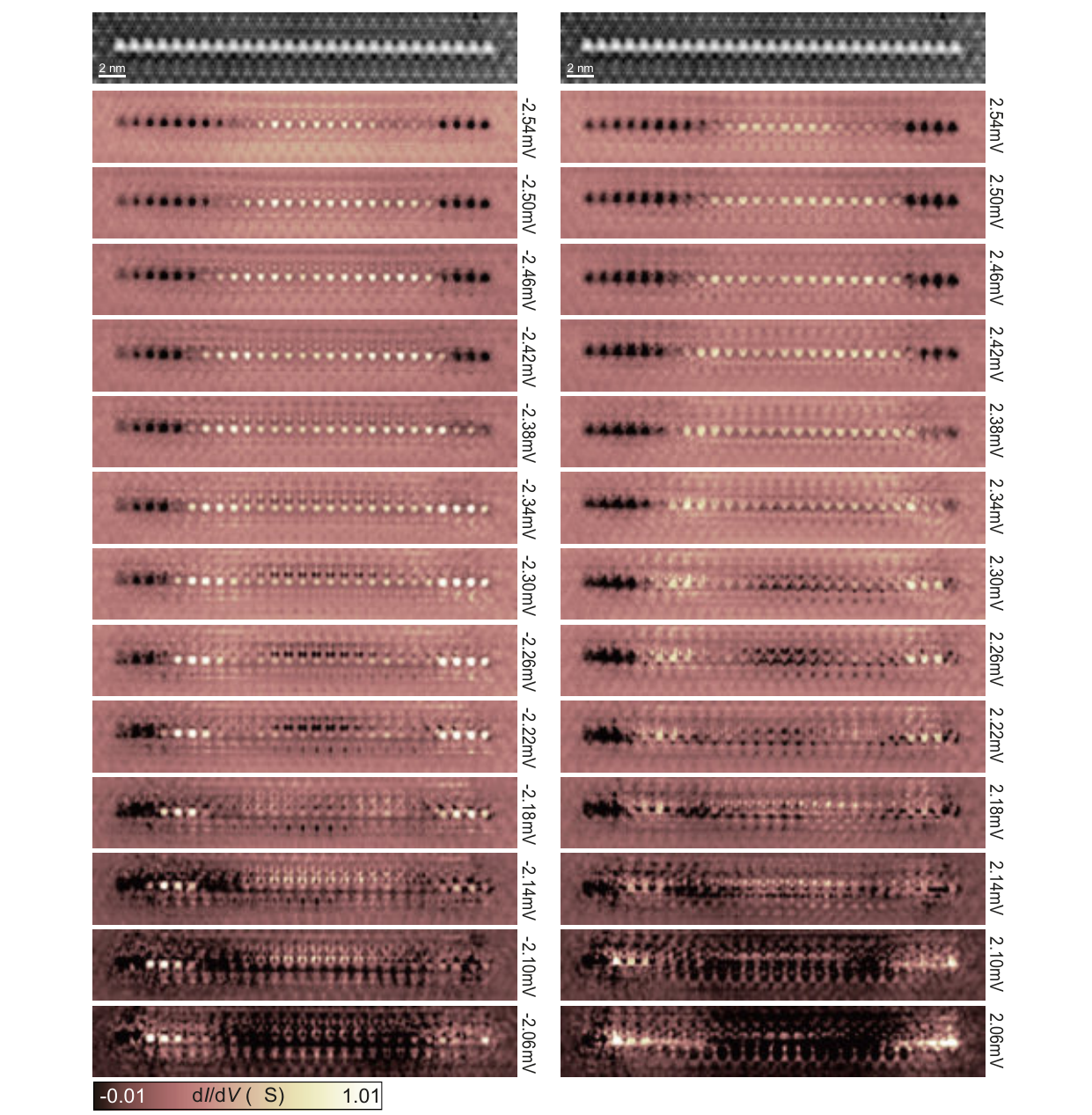}
\caption{\textbf{YSR wave functions of 27-atom chain (part I).} Complementary data to Fig.\,5 in the main part. Constant-contour \didv maps recorded at the voltages given next to each panel ($\Delta_{\mathrm{tip}}\approx\SI{1.55}{\milli\electronvolt}$). Feedback was opened at $\SI{700}{\pico\ampere},\SI{5}{\milli\volt}$ and a modulation of $\SI{15}{\micro\volt}$ was used. Topography can be found in the top (set point: $\SI{100}{\pico\ampere},\SI{10}{\milli\volt}$).
}
\label{Fig:27atoms1}
\end{figure}

\begin{figure}[ht]\centering
\includegraphics[width=0.95\linewidth]{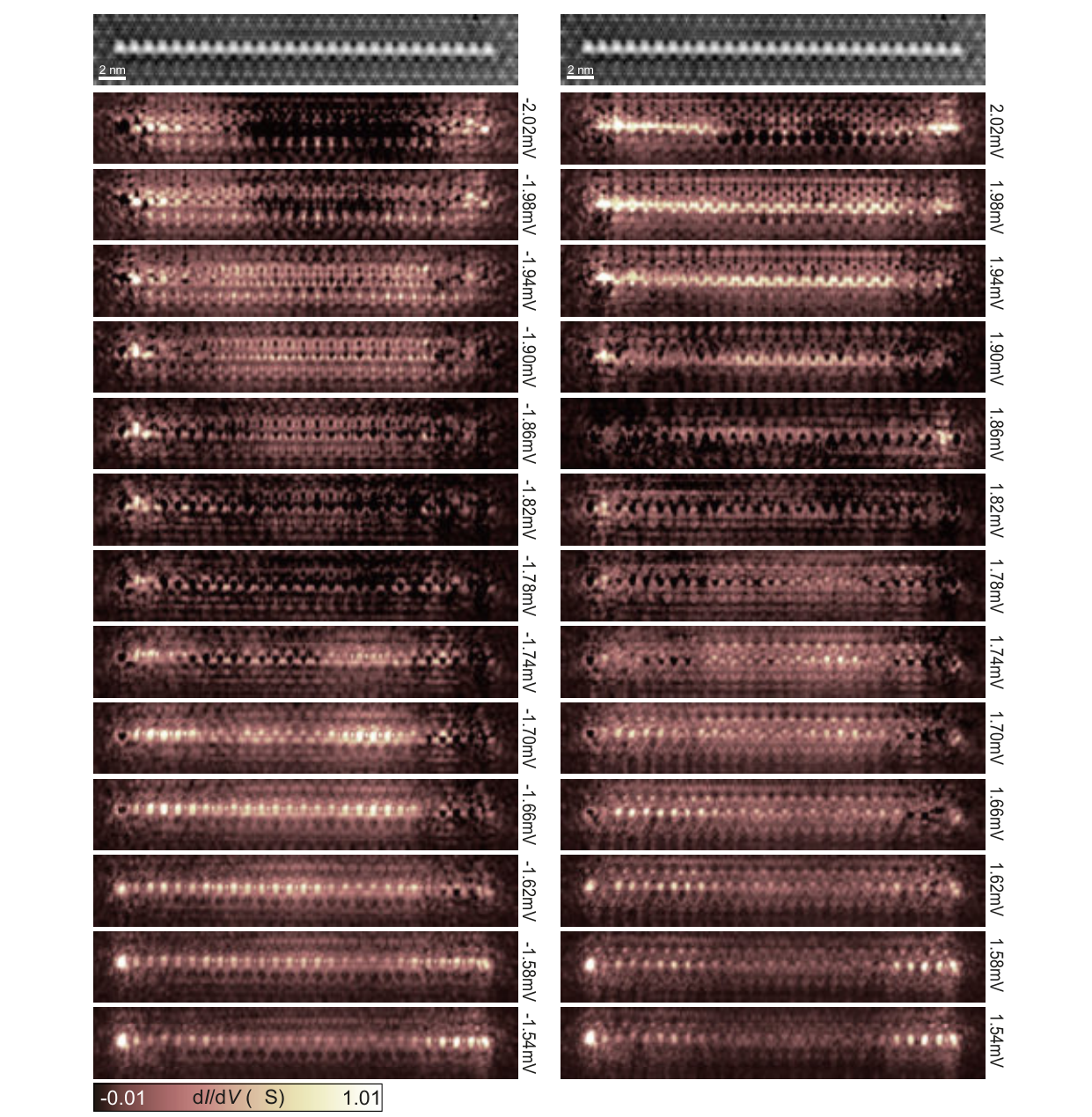}
\caption{\textbf{YSR wave functions of 27-atom chain (part II).} Complementary data to Fig.\,5 in the main part. Constant-contour \didv maps recorded at the voltages given next to each panel ($\Delta_{\mathrm{tip}}\approx\SI{1.55}{\milli\electronvolt}$). Feedback was opened at $\SI{700}{\pico\ampere},\SI{5}{\milli\volt}$ and a modulation of $\SI{15}{\micro\volt}$ was used. Topography can be found in the top (set point: $\SI{100}{\pico\ampere},\SI{10}{\milli\volt}$).
}
\label{Fig:27atoms2}
\end{figure}

In the main text, we discussed the band bending in a 27-atom chain due to the CDW. Supplementary Figures \ref{Fig:27atoms1} and \ref{Fig:27atoms2} show a set of \didv maps, reflecting the energetic and spatial evolution of the YSR bands. These allow to track the evolution of the bands along the chain. For instance, the highest-energy feature in the chain appears at $\SI{2.50}{\milli\volt}$ in the center of the chain, but at $\SI{2.26}{\milli\volt}$ at the chain's terminations. The van Hove singularity of the $\beta$-derived band at $\sim\SI{1.95}{\milli\volt}$ appears at the same energy almost along the entire chain (map at $\SI{1.94}{\milli\volt}$). Similarly, the van Hove singularity of the $\alpha$-derived band at $\SI{-1.70}{\milli\volt}$ exhibits a small shift towards zero energy only at the terminations (see map at $\SI{1.54}{\milli\volt}$).


\clearpage

\begin{figure}[ht]\centering
\includegraphics[width=0.95\linewidth]{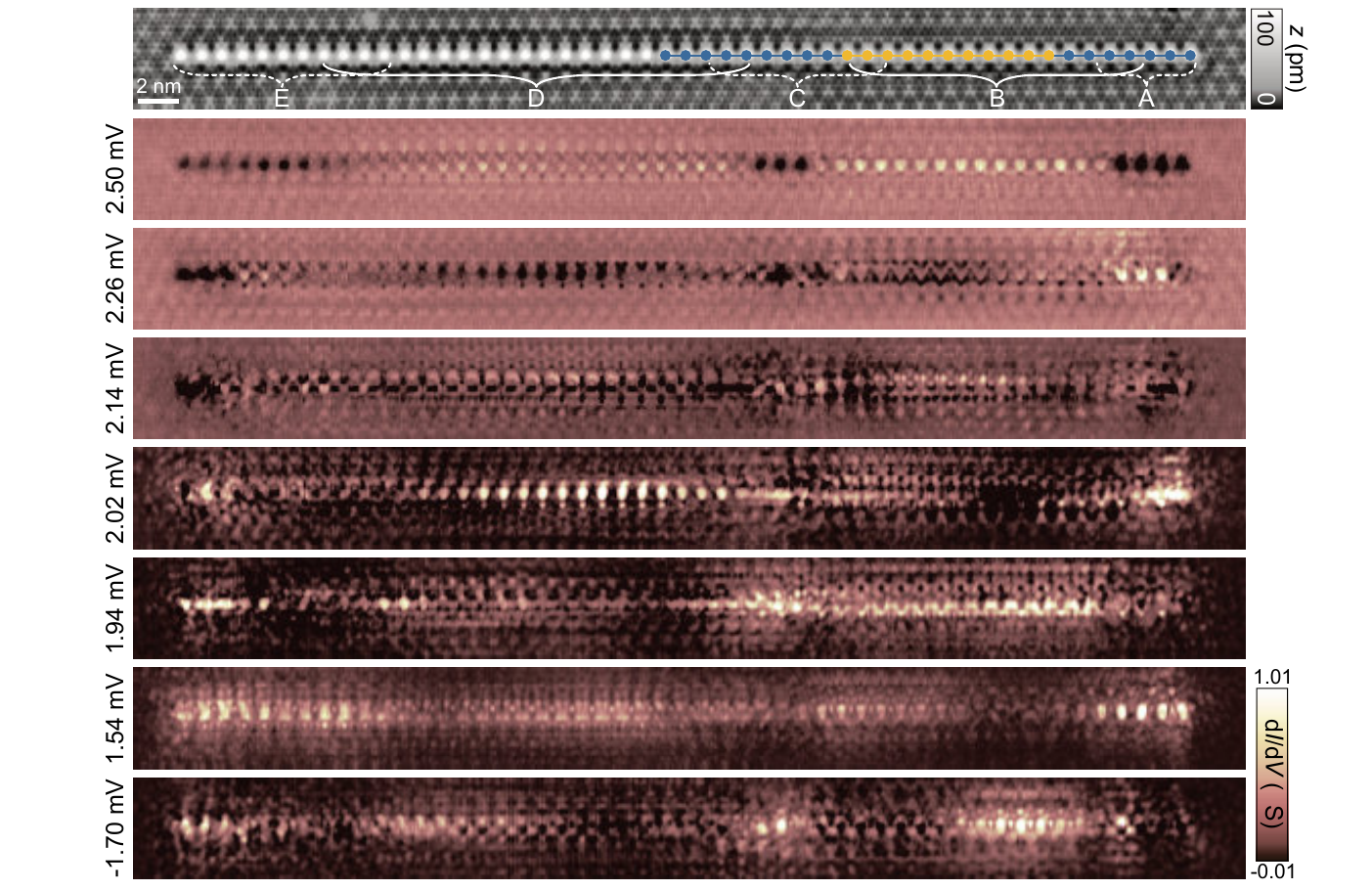}
\caption{ \textbf{YSR wave functions of 51-atom chain across different domains of the CDW.} Selected constant-contour \didv maps recorded on the 51-atom chain (Fig.\,6 in the main part) at the voltages given at the left to each panel ($\Delta_{\mathrm{tip}}\approx\SI{1.55}{\milli\electronvolt}$). Feedback was opened at $\SI{700}{\pico\ampere},\SI{5}{\milli\volt}$ and a modulation of $\SI{15}{\micro\volt}$ was used. Topography can be found in the top (set point: \sptop). The former 11- (27-) atom chain is indicated in yellow (blue). 
}
\label{Fig:51atoms}
\end{figure}

Supplementary Figure\,\ref{Fig:51atoms} shows \didv maps of the 51-atom chain at selected energies, which highlight the localization of the YSR bands in the different regions. In particular, region A shows the down-shifted bands at the chain's termination, region B shows the bands of a chain with Fe atoms sitting on CDW maxima, region C represents the abrupt transition between the Fe atoms located on maxima (region B) and minima (region D) of the CDW, region E shows again band shifts at the chain termination. 

\begin{figure}[h]\centering
\includegraphics[width=0.95\linewidth]{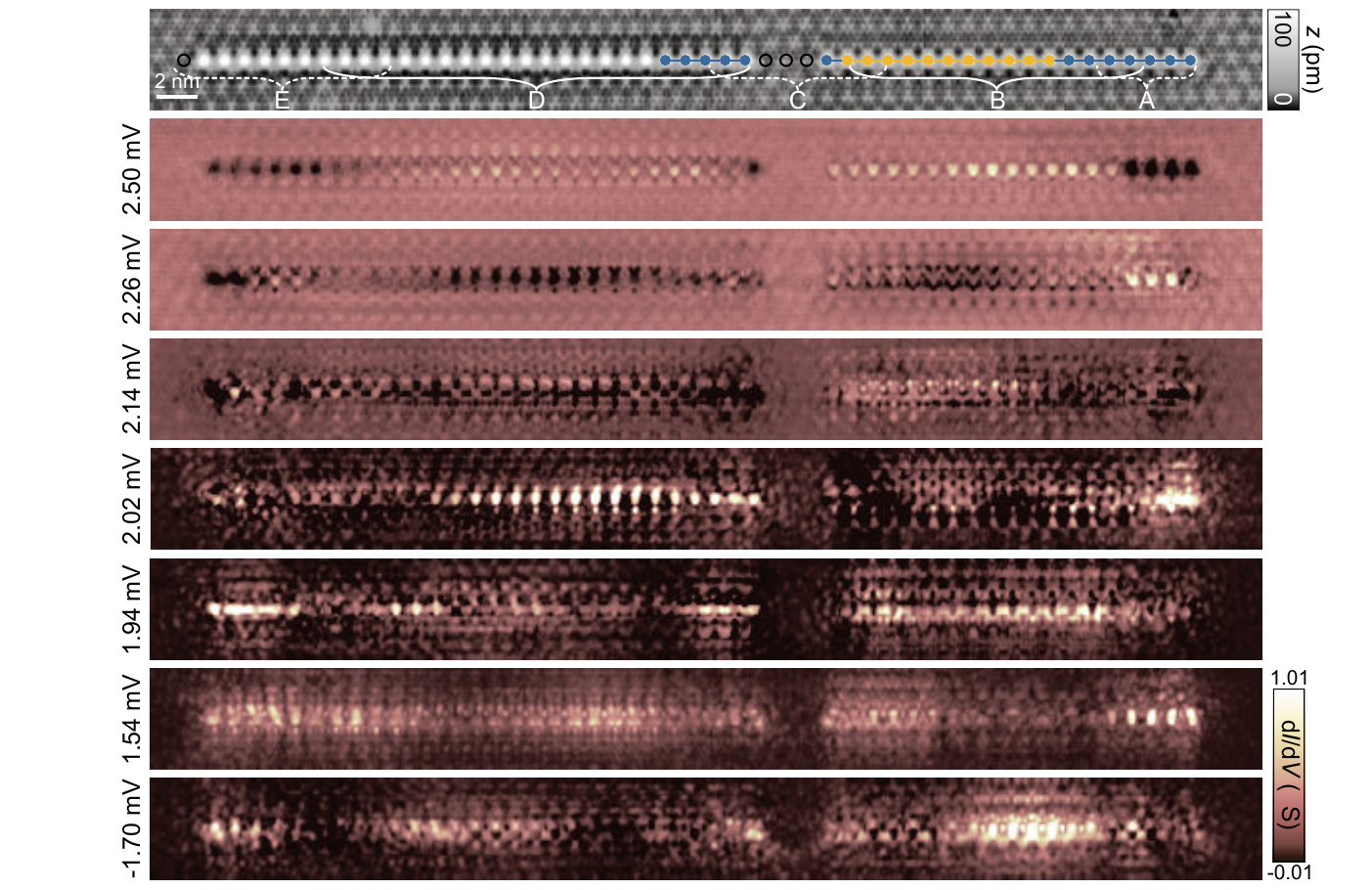}
\caption{\textbf{YSR wave functions of two chains on different domains of the CDW.} Selected constant-contour \didv maps recorded on the former 51-atom chain (Fig.\,6 in the main part) after removal of a few atoms (black circles).  Voltages given on the left to each panel ($\Delta_{\mathrm{tip}}\approx\SI{1.55}{\milli\electronvolt}$). Data in the different sections are very similar to the 51-atom chain. Feedback was opened at $\SI{700}{\pico\ampere},\SI{5}{\milli\volt}$ and a modulation of $\SI{15}{\micro\volt}$ was used. Topography can be found in the top (set point: \sptop).
}
\label{Fig:19-28atoms}
\end{figure}

Removal of three Fe atoms in region C leads to two separate chains (as the CDW starts to distort also on the left end, we also removed one atom there).  \didv maps taken at the same energies as for the uninterrupted chain show very similar intensity distributions (compare Supplementary Fig.\,\ref{Fig:51atoms} and Fig.\,\ref{Fig:19-28atoms}). Regions A, B and D, E of the complete 51-atom chain can thus be interpreted as two non-interacting sub-chains.

\newpage 

\bibliographystyle{apsrev4-2}

\end{document}